\documentclass [preprint,
 amsmath,amssymb,
 aps,
 superscriptaddress,
12pt,floatfix]{revtex4-2}
\usepackage[utf8]{inputenc}
\usepackage{graphicx}
\usepackage{geometry}
\usepackage{xcolor}
\usepackage{amsmath}
\usepackage{placeins}
\usepackage{array}
\usepackage{url}
\usepackage{breakurl}

\renewcommand{\arraystretch}{1.5}

\begin{document}

\title{COVID-19 in low-tolerance border quarantine systems:\\ impact of the Delta variant of SARS-CoV-2}

\author{Cameron Zachreson}
\affiliation{School of Computing and Information Systems, The University of Melbourne, Australia}
\author{Freya~M.~Shearer}
\affiliation{Melbourne School of Population and Global Health, The University of Melbourne, Australia}
\author{David~J.~Price}
\affiliation{Melbourne School of Population and Global Health, The University of Melbourne, Australia}
\affiliation{The Peter Doherty Institute for Infection and Immunity, The Royal Melbourne Hospital and The University of Melbourne, Australia}
\author{Michael~J.~Lydeamore}
\affiliation{Department of Infectious Diseases, The Alfred and Central Clinical School, Monash University, Australia}
\author{Jodie~McVernon}
\affiliation{Melbourne School of Population and Global Health, The University of Melbourne, Australia}
\affiliation{The Peter Doherty Institute for Infection and Immunity, The Royal Melbourne Hospital and The University of Melbourne, Australia}
\author{James McCaw}
\affiliation{Melbourne School of Population and Global Health, The University of Melbourne, Australia}
\affiliation{The Peter Doherty Institute for Infection and Immunity, The Royal Melbourne Hospital and The University of Melbourne, Australia}
\affiliation{School of Mathematics and Statistics, The University of Melbourne, Australia}
\author{Nicholas~Geard}
\affiliation{School of Computing and Information Systems, The University of Melbourne, Australia}
\affiliation{The Peter Doherty Institute for Infection and Immunity, The Royal Melbourne Hospital and The University of Melbourne, Australia}

\date{\today}

\begin{abstract}
In controlling transmission of COVID-19, the effectiveness of border quarantine strategies is a key concern for jurisdictions in which the local prevalence of disease and immunity is low. In settings like this such as China, Australia, and New Zealand, rare outbreak events can lead to escalating epidemics and trigger the imposition of large scale lockdown policies. Here, we examine to what degree vaccination status of incoming arrivals and the quarantine workforce can allow relaxation of quarantine requirements. To do so, we develop and apply a detailed model of COVID-19 disease progression and transmission taking into account nuanced timing factors. Key among these are disease incubation periods and the progression of infection detectability during incubation. Using the disease characteristics associated with the ancestral lineage of SARS-CoV-2 to benchmark the level of acceptable risk, we examine the performance of the border quarantine system for vaccinated arrivals. We examine disease transmission and vaccine efficacy parameters over a wide range, covering plausible values for the Delta variant currently circulating globally. Our results indicate a threshold in outbreak potential as a function of vaccine efficacy, with the time until an outbreak increasing by up to two orders of magnitude as vaccine efficacy against transmission increases from 70\% to 90\%. For parameters corresponding to the Delta variant, vaccination is able to maintain the capacity of quarantine systems to reduce case importation and outbreak risk, by counteracting the pathogen's increased infectiousness. To prevent outbreaks, heightened vaccination in border quarantine systems must be combined with mass vaccination. The ultimate success of these programs will depend sensitively on the efficacy of vaccines against viral transmission.

\end{abstract}

\maketitle

\section{Introduction}

Mitigation of pandemics requires a continuous analysis of risk trade-offs in order to respond proportionately and efficiently. Border quarantine systems are designed to allow travel between jurisdictions while limiting the risk of disease transmission between them. Rigorously limiting disease transmission between regions is appropriate when large differences exist in pathogen prevalence.  While strict border measures are effective at preventing disease incursions, they are also costly to operate and reduce international travel to a trickle. Given the enormous economic and social costs associated with the international travel restrictions that come with stringent border quarantine policies, such systems should only be used when they can prevent a catastrophic public health crisis \cite{guan2020global,world2021policy}.  

During the COVID-19 pandemic, border quarantine strategies have been implemented in most parts of the world in various forms \cite{wells2020impact,haug2020ranking}. In regions with tightly controlled borders, screening of international travellers has provided an effective means of limiting the importation rate of individuals infected with SARS-CoV-2 \cite{steyn2021managing,gostic2020estimated,ashcroft2021quantifying}. This has facilitated the success of outbreak control strategies relying on targeted test-trace-isolate-quarantine (TTIQ) responses. This combination of approaches has been largely successful in preventing widespread epidemics of COVID-19 in countries such as China, New Zealand, and Australia \cite{lau2020positive,zachreson2021risk,baker2020new,summers2020potential}. In Australia, this has meant that many citizens abroad at the outset of the pandemic have been stranded overseas due to quarantine capacity constraints. It has also stressed the higher education and tourism sectors and devastated the airline industry \cite{thatcher2020predicting,tisdall2021covid,beck2020insights}. The design of a quarantine system needs to balance the benefits of reducing the risk of importation against the associated costs. In order to assess this trade-off, analytical frameworks must incorporate emerging evidence about pathogen characteristics to provide accurate estimates of the risk associated with alternative quarantine strategies.

Two potential factors motivating a revaluation of quarantine stringency include changes in the properties of a pathogen, and changes in what is deemed an acceptable level of breach risk. In the case of COVID-19, the development and rollout of effective vaccines provided an opportunity for countries that had previously maintained stringent border controls to contemplate a future in which these measures could be relaxed. 

However, the emergence of the Delta variant of SARS-CoV-2 has produced the need to evaluate risk trade-offs in the context of a virus exhibiting higher transmissibility, higher clinical severity, and against which existing vaccines are less effective \cite{fisman2021progressive,zhang2021transmission,kang2021transmission,dagpunar2021interim,brown2021outbreak,chia2021virological,lopez2021effectiveness,elliott2021react}. In this shifting context, the role of vaccination and the acceptable risk of border quarantine breach events must be re-evaluated. The emergence of new variants with broadly different disease characteristics can be viewed as the onset of a new pandemic. Therefore, risk trade-offs and mitigation measures need to be assessed based on up-to-date information. In this work, we evaluate the efficacy of border quarantine systems as a function of the following pathogen characteristics:
\begin{itemize}
    \item{transmissibility}
    \item{efficacy of vaccines against transmission (the combination of efficacy against infection, and efficacy against transmission from breakthrough cases)}
\end{itemize}
As our primary purpose is to estimate reduction in transmission risk, we focus on pathogen and vaccine characteristics associated with transmission. The efficacy of vaccines for preventing infection and onward transmission is a primary consideration when designing modified border quarantine pathways for vaccinated travellers, which is currently becoming a widely adopted framework \cite{osama2021covid}.

We evaluate the performance of a simulated border quarantine system consistent with the recommendations of the World Health Organization (WHO) for countries that choose to quarantine all international arrivals \cite{world2021considerations}. This includes a 14-day minimum stay, a testing regime, and a response strategy that isolates confirmed cases and their contacts from the other travellers in quarantine (Figure \ref{fig:schematic_system} and Figure \ref{fig:schematic_response}). The chosen model reproduces general features adopted by the Chinese, Australian, and New Zealand border quarantine systems. 

We examine the performance of the quarantine system over a range of vaccine efficacy levels and reproductive ratios, relative to an unvaccinated baseline condition (0\% vaccine efficacy). We then investigate how the risk of outbreaks seeded by quarantine breach events changes with the emergence of a more transmissible strain (i.e., the Delta variant of SARS-CoV-2). For our outbreak scenarios, we examine the effect of varying levels of population-wide vaccination coverage. This model, and the ensemble of results presented here can help guide adaptation of border quarantine measures as the virus evolves, vaccination coverage increases, and vaccine efficacy changes. 

\section{Methods}
\subsection{Model overview (structure, inputs, outputs)}

\begin{figure}
    \centering
    \includegraphics[width = \textwidth]{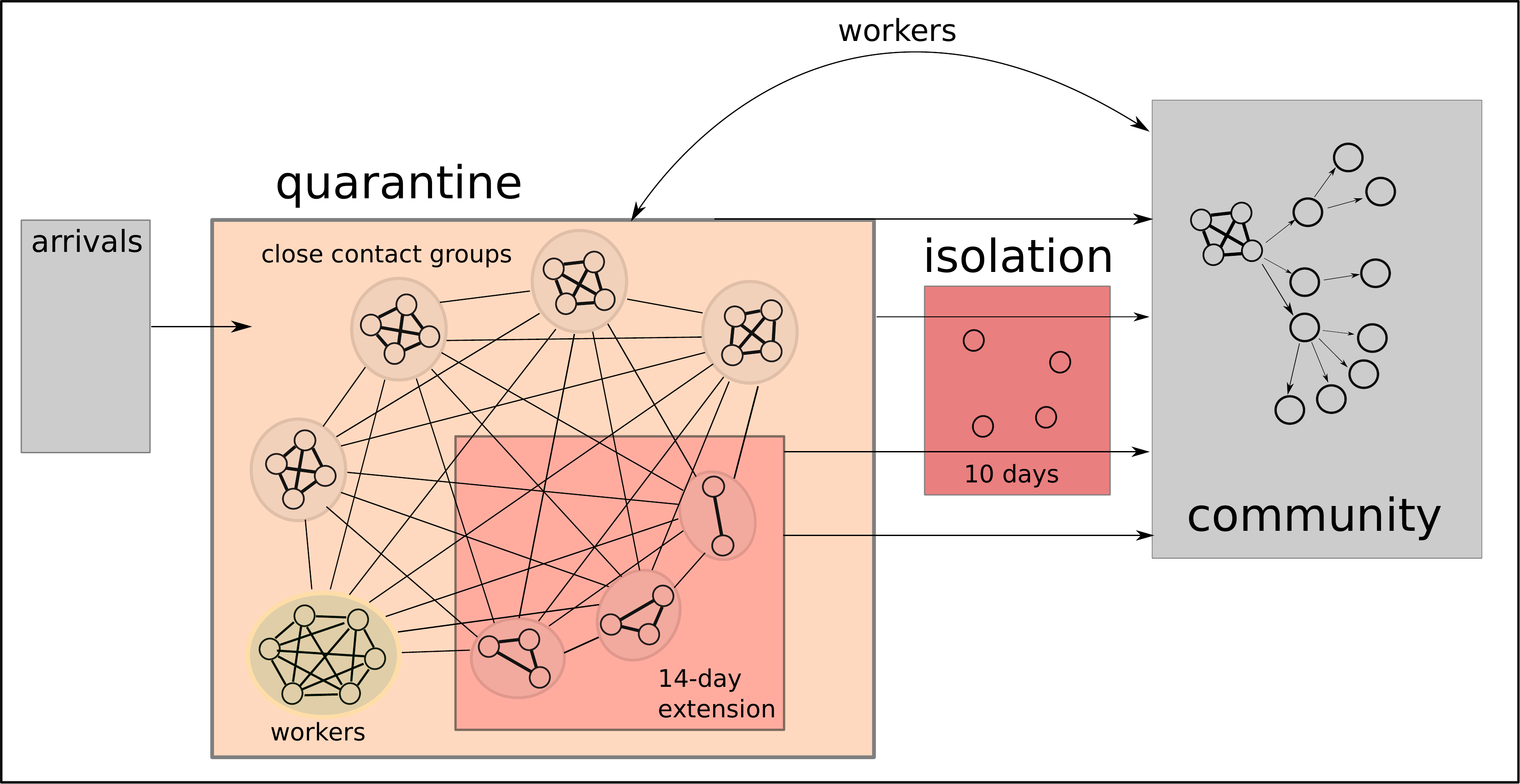}
    \caption{Schematic of the quarantine system model. Arrivals enter quarantine in groups of four close contacts. These groups are in weak contact with one another and with the workforce. If a case is detected, the infected individual is placed into 10-day isolation, and their contacts are placed into a 14-day quarantine extension. Travellers in extended quarantine are still in contact with the workforce and with other close-contact groups within the facility. A separate branching process model is used to evaluate the potential for outbreaks in the community based on the breach event statistics produced by the quarantine system model.}
    \label{fig:schematic_system}
\end{figure}

We simulate virus transmission, case detection, and isolation response pathways within a single quarantine facility with a capacity of 100 travellers. The facility is staffed by 20 vaccinated workers who have intermittent contact with those in quarantine. Individuals are processed subject to testing and case isolation. The structure of the quarantine environment is generic, but captures the main principles typically applied in border quarantine facilities. The output of the quarantine model is used in a separate branching process model to evaluate the potential for outbreaks of community transmission. A schematic of the overall system is shown in Figure \ref{fig:schematic_system}.

\begin{figure}
    \centering
    \includegraphics[width = \textwidth]{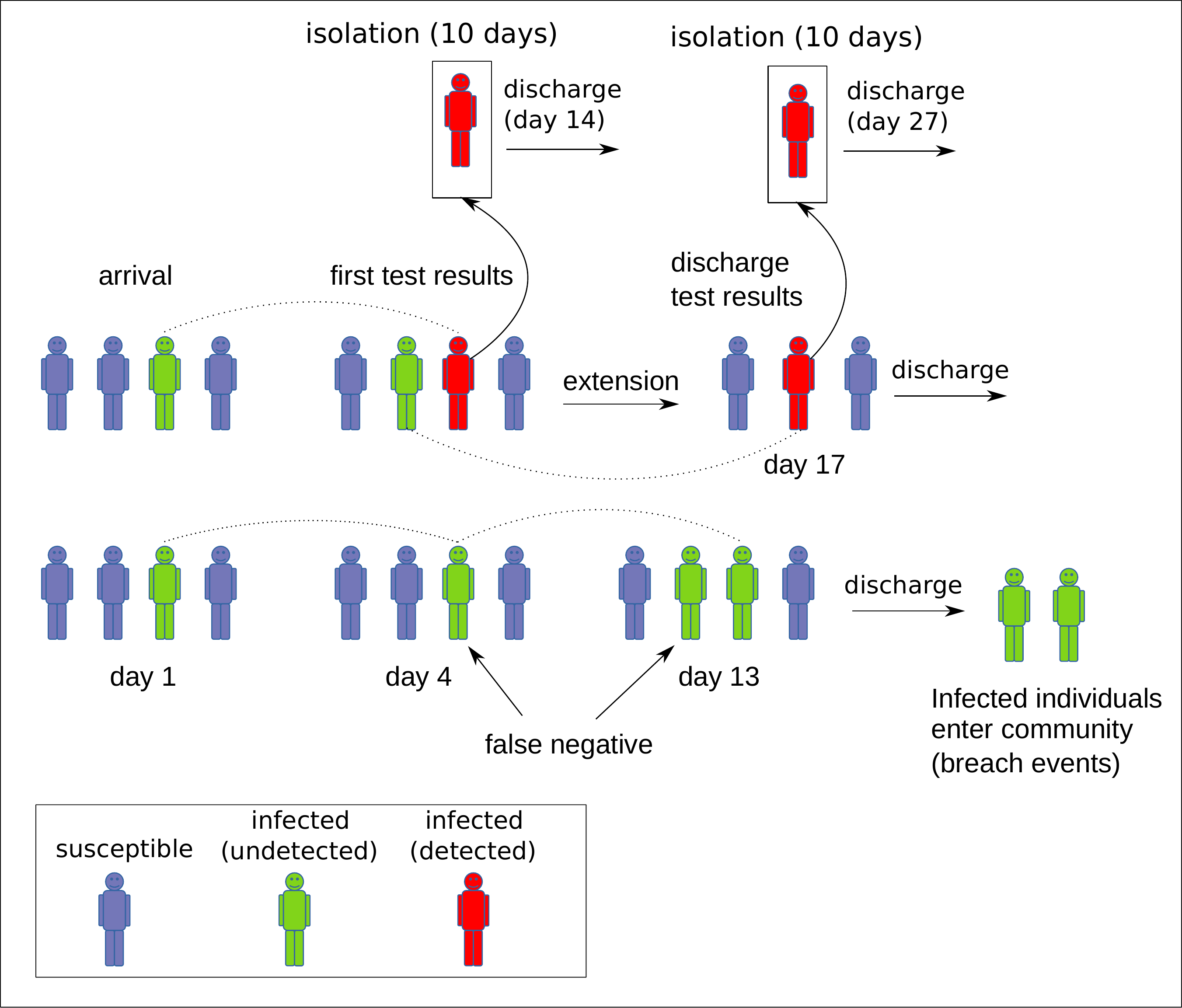}
    \caption{Schematic of event sequence from arrivals entering quarantine, through testing, isolation of detected infections, extension of quarantine for close contacts (i.e., family groups), and eventual discharge from quarantine. False negative discharge tests lead to infected individuals entering the community.}
    \label{fig:schematic_response}
\end{figure}

The population is structured in two distinct groups, one of these comprising workers who staff the facility, and the other comprising the quarantined travellers. Travellers move through the system as indicated in Figure \ref{fig:schematic_response}. After arriving in close contact groups of 4 individuals, travellers remain in the system for 14 days unless an infection is detected within their group. The 14 day minimum stay means that, typically, 50 travellers exit the system each week. 

The model used to simulate border quarantine incorporates a detailed description of COVID-19 progression and transmission that captures the following salient features:
\begin{itemize}
    \item {A lognormally distributed incubation period (mean approx. 5.5 d $[\mu = 1.62, \sigma = 0.418]$, Figure \ref{fig:tinc_compare})}
    \item {Time-varying infectiousness, increasing from the moment of exposure, peaking just before symptom onset, and declining until recovery (Figure \ref{fig:beta_vs_t})}
    \item {Time-varying RT-PCR test sensitivity with a peak before symptom onset followed by a gradual decline (Figure \ref{fig:tsens_i_traj})}
    \item {An over-dispersed secondary case distribution (Figure \ref{fig:nbinom})}
\end{itemize}
These features allow the model to capture two important effects of the quarantine environment. The first of these is the truncation of the naturally over-dispersed secondary case distribution due to physical separation of close contact groups. The second is the tendency for false negative tests to occur during the early stages of infection. Detailed descriptions of disease natural history and test sensitivity models can be found in the Supporting Information.

Detection of infection can occur due either to positive RT-PCR tests (conducted on days 3 and 12), or symptom onset. The model assumes that 1/3 of all cases are asymptomatic. Asymptomatic cases can only be detected through testing. Symptomatic cases may be detected during the pre-symptomatic period through RT-PCR. If not, they are detected at the end of their incubation period, when they begin expressing symptoms (all symptomatic individuals are treated as confirmed cases). 

Case detection in travellers results in a 10-day isolation period for the case, as well as a 14-day quarantine extension for close contacts with additional tests on days 3 and 12 of the extension period. Subsequent detection within the same group of contacts results in isolation of cases, but does not incur additional extensions for remaining contacts. Therefore, the maximum period for which an individual can remain in the system is 38 days. This would occur if an individual were the close contact of a case detected on day 14 of the initial stay, were detected as a case themselves on the 14th day of their extension period, and were subsequently isolated for 10 additional days. While in quarantine extension, transmission dynamics are not altered. While in isolation, individuals may not transmit infection to any other individuals. When all members of a close contact group are discharged from the system, they are replaced by a new group.

In a simulated quarantine facility, the workforce is composed of 20 individuals who come and go each day. Workers are tested via RT-PCR on each day they attend the site. Each worker attends for 5 days per week and has two days off per week. Workers may become infected through contact with quarantined travellers, or through contact with infected co-workers. The force of infection applied between travellers and workers is reduced by a factor of 100 to simulate infection control measures (e.g., mask wearing), and limited contact. On the other hand, the force of infection between workers is only reduced by a factor of 10 relative to unmitigated contact. This accounts for infection control, with higher levels of mixing. Because workers are tested frequently, infections are typically detected during the pre-symptomatic period. Infected workers are replaced with susceptible ones after either detection or recovery. We note that replacement after recovery is not strictly realistic but avoids eventual saturation of the recovered worker population over long simulations.

\subsection{Scenarios: vaccine efficacy and pathogen transmissibility}

The scenarios we selected are designed to determine how the risk associated with quarantine breaches is mitigated within a pathway exclusive to fully vaccinated travellers. Such a pathway would be staffed exclusively by vaccinated workers as well, for 100\% vaccination coverage within the system. In this context, we examine performance over a wide range of vaccine efficacy and viral transmissibility parameters. Performance is determined relative to a baseline scenario, in which vaccine efficacy ($VE$) is set to 0, and disease transmission parameters are aligned to COVID-19 strains in circulation prior to the emergence of the Delta variant. 

In the quarantine and community transmission model systems, vaccines play three important roles: 
\begin{itemize}
    \item vaccination of travellers prior to departure reduces the proportion of infected arrivals by a factor of $(1 - VE)$ from the base rate of 1\%
    \item vaccination of workers and travellers limits transmission within quarantine (which reduces the rate of breach events)
    \item mass vaccination prevents outbreaks in the community when quarantine breach events occur
\end{itemize}

For efficacy of vaccination against transmission, we investigate the range between 0\% and 90\% total efficacy. While the quarantine model treats vaccine efficacy as a combination of efficacy against infection $V_I$ and efficacy against onward transmission from breakthrough infections $V_T$, we simplify to a single efficacy parameter for the scenarios investigated here. This is because we assume all individuals in the system are vaccinated. We note that for scenarios in which only a subset of individuals are vaccinated, this simplifying assumption would need to be relaxed to account for interactions involving combinations of vaccinated and unvaccinated individuals. As noted above, we consider efficacy of 0\% as a baseline, accounting for the levels of incursion risk existing before vaccines became available. Efficacy levels of 80\%-90\% are indicative of the conditions existing for the COVID-19 ancestral lineage and Alpha variants previously in circulation \cite{nasreen2021effectiveness,zachreson2021how}. Lower efficacy ranges can account for the emergence of variants capable of higher levels of breakthrough infection (such as the currently dominant Delta variant). 

The emergence of viral variants also requires us to investigate a broad range of transmission rates. The transmissibility of the Delta variant has been estimated to be approximately twice that of the ancestral (non VOC) lineages \cite{campbell2021increased}. Based on a well-traced outbreak in China (Guangdong province, May 2021), the basic reproductive ratio for the Delta variant is estimated to be approximately 6 \cite{kang2021transmission,zhang2021transmission}. Therefore, in order to understand the scaling of quarantine system performance with disease transmissibility, we examine a wide range of possibilities from $R_0 = 1$ to $R_0 = 10$.

\subsection{Model variations: shorter incubation period}
In our baseline model, we sample the viral incubation period (time between infection and symptom onset) as described in studies of the ancestral lineage \cite{lauer2020incubation}. This estimate also influences our model of test sensitivity as a function of time from symptom onset \cite{hellewell2021estimating} (see Supporting Information). 

Quarantine systems operating under the ``test and release" framework are primarily designed to prevent individuals infected elsewhere from entering the community. The timing considerations applied to this operational framework (i.e., a 14-day minimum stay) are designed to exceed the disease incubation period. From this perspective, the emergence of variants with incubation periods different to those for which the system was designed, can be expected to alter system performance. System performance should, in general, improve for shorter incubation periods. 

In the context of COVID-19, recent reports suggest that the Delta variant may have a shorter incubation period than that of the ancestral lineage \cite{zhang2021transmission}. However, other reports using data from the same outbreak indicate that the incubation period of the Delta variant has not changed substantially \cite{kang2021transmission}. Given the preliminary evidence for a shorter incubation period, we performed a sensitivity analysis of this key parameter (see Supporting Information).

\subsection{Model outputs}

The quarantine facility model produces a timeseries of ``breach events" each of which corresponds to an infected individual (worker or traveller) interacting with the community outside of quarantine. For an infected traveller, this can occur for two reasons: (1) leaving case isolation while still infectious (i.e., after the 10-day isolation period), or (2) leaving quarantine after a false negative test result. The recorded breach event accounts for the number of days over which the individual will remain infectious after leaving quarantine, and the integrated force of infection produced by the individual over that period. For infected workers, recorded breach events account for the time from infection to either detection or recovery, and the integrated force of infection over that period. In general, the community force of infection produced by a breach event associated with case $i$ is given as:
\begin{equation}
    \beta_i = \sum_{t~\in~t_c} \beta(t, i) \Delta t \,, 
\end{equation}
where $\beta_i$ is the integrated force of infection produced by agent $i$ outside of quarantine, $t_c$ represents the set of discrete timepoints over which individual $i$ is infectious in the community, $\beta(t, i)$ is the time-dependent force of infection for case $i$, and $\Delta t$ is the discrete time step used in the simulation (here, $\Delta t = 0.1~\text{days}$). 

\begin{figure}
    \centering
    \includegraphics[width = \textwidth]{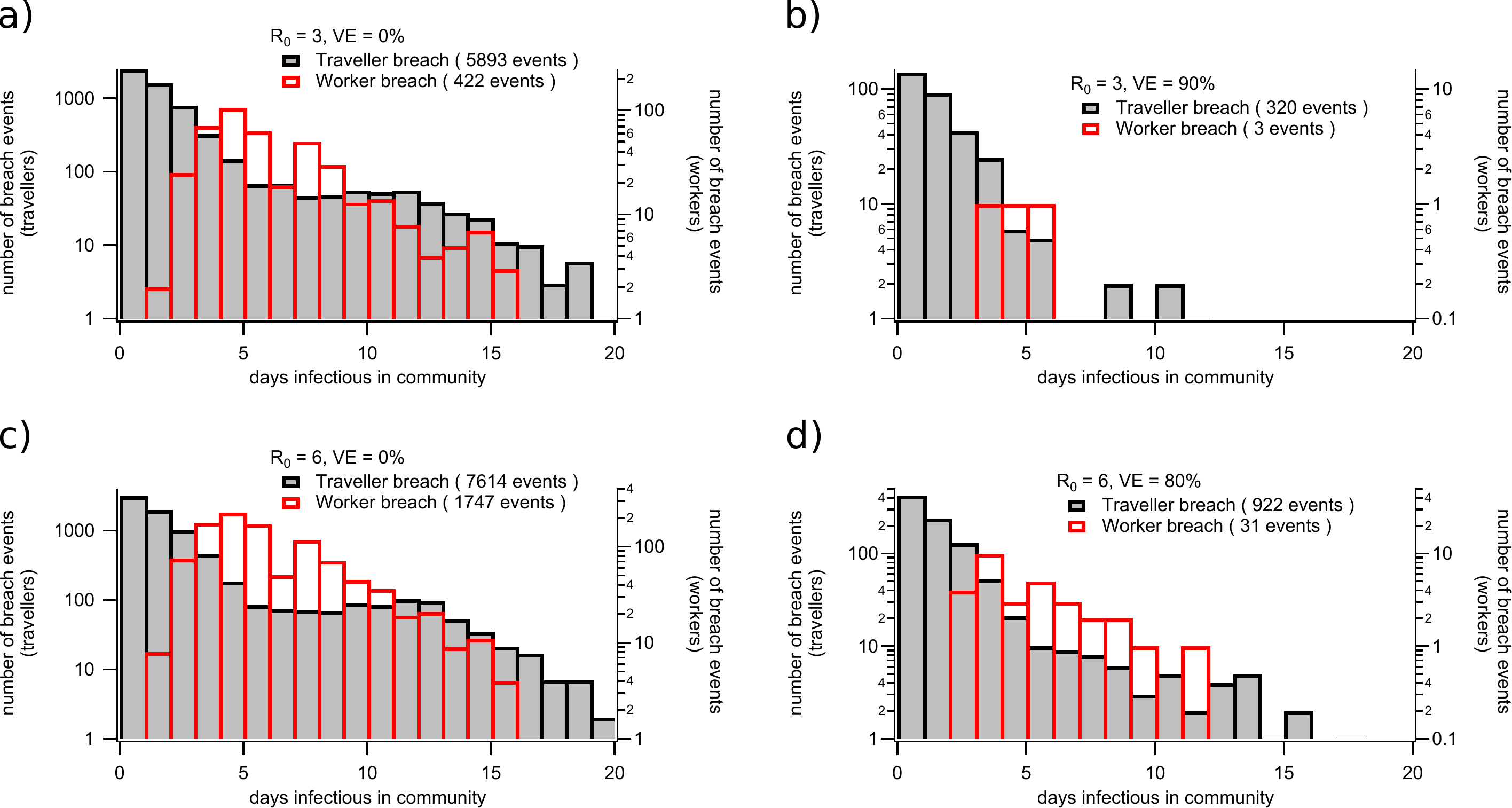}
    \caption{Distribution of the days infectious in the community produced by each breach event simulated in four illustrative scenarios: (a) ancestral strain without vaccination, (b) ancestral strain with vaccination, (c) Delta strain without vaccination, (d) Delta strain with vaccination (increased transmissibility, decreased vaccine efficacy). Breach events can occur from infected travellers being discharged from quarantine while still infectious. They can also occur or due to infected workers that are not detected while pre-symptomatic or asymptomatic due to false-negative RT-PCR screening tests. The distribution of days infectious in the community differs qualitatively for traveller or worker-related breach events. Each subfigure corresponds to a different combination of reproductive ratio $R_0$ and vaccine efficacy $VE$. Shaded black bars correspond to traveller-related breach events while open red bars correspond to worker-related breach events. }
    \label{fig:eday_hist}
\end{figure}

Breach events are rare due to the effectiveness of the quarantine system, particularly when those within it are vaccinated at high efficacy. To generate a large number of breach events for use in comparing outbreak statistics between scenarios, each simulation lasts for $10^6~\text{days}$. Recall that workers are tested daily so that infections are typically detected during the pre-symptomatic period. Infected workers are replaced after either detection or recovery (the latter avoids eventual saturation of the recovered worker population over long simulations). The different conditions for breach events involving travellers and workers produce qualitatively different breach statistics that depend also on vaccine efficacy and $R_0$ (Figure \ref{fig:eday_hist}). More details of the quarantine simulation model can be found in the Supporting Information. 

We evaluated the effectiveness of the quarantine system by examining the integrated force of infection introduced into the community due to breach events. This value is computed as:
\begin{equation}
\beta_{tot} = \sum_{i = 1}^{n} \beta_i\,, 
\end{equation}
where $n$ is the total number of breach events simulated in a given scenario, and $\beta_i$ is the force of infection in the community produced by breach $i$. Because all other model parameters are held constant (see Methods), $\beta_{tot}$ is a useful representation of the relative performance of the system for different combinations of $VE$ and $R_0$. In Figure \ref{fig:net_FoI_vs_R0_x_VE}, $\beta_{tot}$ computed for each set of parameters is shown relative to the value computed for the baseline scenario with $R_0 = 3$ and $VE = 0\%$. 

Here, we assume that breakthrough cases are just as contagious as infections in unvaccinated individuals. That is, we assume the vaccine acts primarily to protect those immunised from infection ($V_T = 0,~V_I = VE$). While there is evidence for reduced periods of viral shedding in vaccinated individuals, peak viral loads appear to be similar \cite{riemersma2021vaccinated}. Therefore, we have presented the results based on a conservative assumption, that can be relaxed or modified as new evidence emerges (see the Supporting Information for a sensitivity analysis given an alternate assumption that $V_T = VE$ and $V_I = 0$).  

The influence of the quarantine system on outbreak risk was computed by using the distribution of breach events produced by the quarantine simulation to sample seeding events for the branching process model. In these outbreak scenarios, the level of population-wide vaccination is varied, assuming a negligible level of infection-acquired immunity. The results of the branching process model are used to estimate the probability of a community outbreak, given a fixed volume of travellers and a set infection prevalence. This produces an absolute outbreak risk, that we express as the time until the probability of a transmission cluster containing more than 5 cases exceeds 95\% ($t_{95}$). These values should be interpreted as a means of comparing alternate scenarios, given fixed quantities of incoming arrivals (see the Supporting Information for details of the branching process model).

\section{Results}

\subsection{Summary}

Our results demonstrate how the quarantine system would perform under different conditions of vaccine efficacy and viral transmissibility under fixed assumptions with respect to details such as test schedules and quarantine duration (see Supporting Information). We quantify this in terms of the {\it breach events} produced under each set of conditions. Breach events occur when an infected traveller or quarantine worker comes into contact with the general population. First, we examine the force of infection produced by breach events, which serves as a measure of quarantine system performance independent of the community in which it is embedded. Then, we examine the potential for outbreaks caused by breach events, taking into account the level of vaccination coverage in the general population. 

We also investigate the sensitivity of model outcomes to the disease incubation period, which may be shorter for the Delta variant of COVID-19. The results of this analysis show that shorter incubation periods make the quarantine system more effective because: (i) overall infectious periods are shorter, (ii) test sensitivity increases more quickly after infection, and (iii) the shorter delay to symptom expression hastens detection after false negative arrival tests. 

Finally, we examine the sensitivity of our results to the choice of vaccine efficacy decomposition. We treat disease transmission as a two-part process first requiring infection (subject to the efficacy term $V_I$ {\it efficacy against infection}, and then requiring onward transmission (subject to the efficacy term $V_T$ {\it efficacy against onward transmission}). In this model the overall efficacy against transmission $VE$ is given as 
\begin{equation}\label{eq:VE}
VE = 1 - (1 - V_T)(1 - V_I) \,,
\end{equation}
which means that two extreme interpretations exist: (1) $VE = V_I$, $V_T = 0$ and (2) $V_T = VE$, $V_I = 0$. Our main results use the first extreme (see the Methods section for a more detailed discussion of this choice). We investigate the second extreme in our sensitivity analysis. This gives the range of results over which any decomposition consistent with Equation \ref{eq:VE} may fall.

\subsection{Quarantine system breach risk}

\begin{figure}
    \centering
    \includegraphics[width = \textwidth]{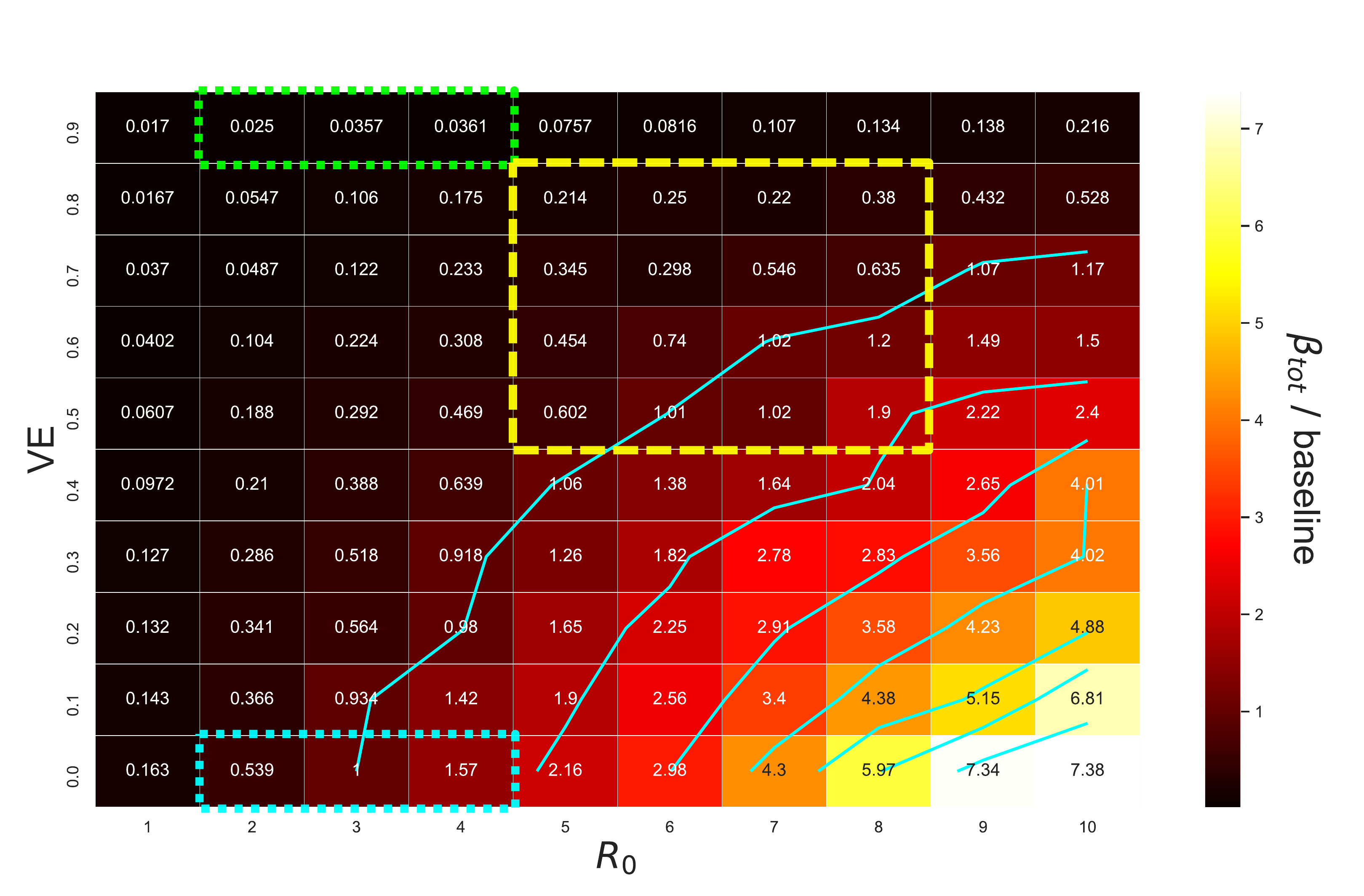}
    \caption{Integrated force of infection relative to baseline computed from quarantine breach events simulated by the model. The heatmap and contour demonstrates how this value scales with vaccine efficacy ($VE$), and the basic reproductive ratio of the virus ($R_0$). In these simulations, all incoming arrivals and quarantine workers are vaccinated, with susceptibility to infection reduced by the factor indicated by $VE$ (i.e., $VE = V_I$, $V_T = 0$). The dotted blue box represents plausible values for the baseline condition when the ancestral lineage and Alpha variant of SARS-CoV-2 were dominant and no vaccines were available. The green dotted box represents the scenarios corresponding to vaccinated quarantine pathways before emergence of the Delta variant. The yellow dashed box covers a range of values plausible for Delta variant scenarios.}
    \label{fig:net_FoI_vs_R0_x_VE}
\end{figure}

Interpreting $\beta_{tot}$ as a measure of outbreak potential under each set of conditions, the heatmap in Figure \ref{fig:net_FoI_vs_R0_x_VE} illustrates how vaccine efficacy must increase to offset the rise in outbreak potential produced by increases in $R_0$. From baseline conditions ($VE = 0\%$, $R_0 = 3$), following the outermost contour delineated in Figure \ref{fig:net_FoI_vs_R0_x_VE} illustrates that vaccine efficacy must exceed 60\% in order for baseline risk levels to be maintained for an $R_0$ of 6, and must exceed 70\% for $R_0 = 8$. The required $VE$ levels saturate for high values of $R_0$, with efficacy of 70\% to 80\% sufficient to maintain baseline risk levels even for $R_0 = 10$. This saturation occurs because transmission within the quarantine environment is partially constrained by the grouping of arrivals into small cohorts (i.e., family units). This constraint on transmission is contingent upon substantial reduction of exposure risk outside of close contact groups, representative of stringent infection control measures within a facility. 

\subsection{Community outbreak risk}

Next we simulate outbreaks based on the distribution of quarantine breach events produced by the quarantine model. For these scenarios we investigate a subset of $VE$ and $R_0$ combinations corresponding to plausible values for the Delta variant of COVID-19. We investigated population outbreak characteristics for $VE$ $\in [0.5, 0.6, 0.7, 0.8, 0.9]$ and for $R_0 \in [5, 6, 7, 8, 9, 10]$, while varying mass vaccination coverage levels. For the branching process scenarios, a fixed traveller arrival rate of 3000/wk was assumed, with an infection prevalence of 10\%. This differed from the inflow assumptions of the quarantine model (50 travellers per week, 1\% infection prevalence), so a scaling factor was used to linearly adjust the breach rate produced by the quarantine model (see Methods in Supporting Information). This scaling approximation assumes linear dependence of the quarantine breach rate with incoming arrival infection prevalence up to 10\%.

\begin{figure}
    \centering
    \includegraphics[width =0.8 \textwidth]{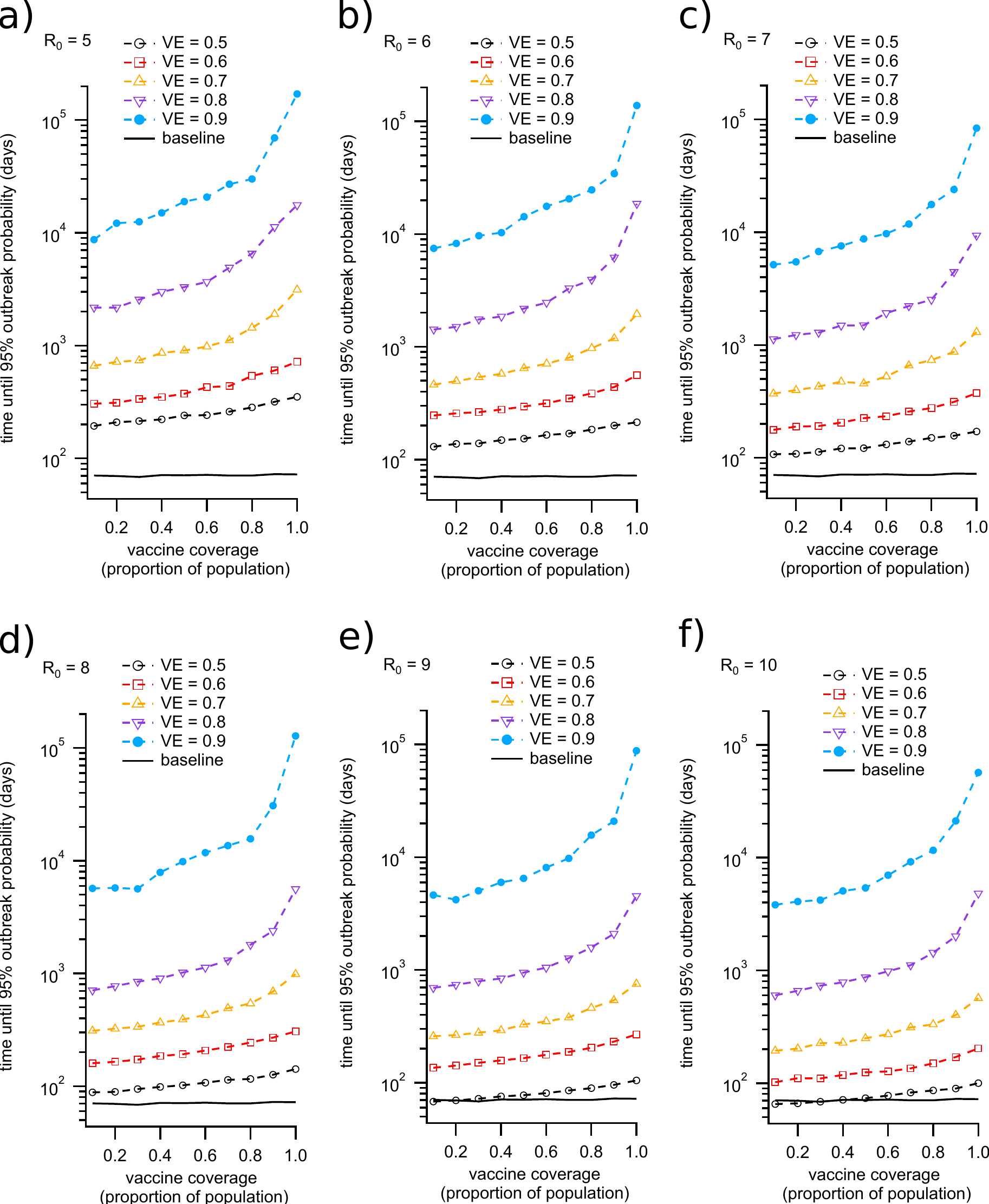}
    \caption{Time until probability of an outbreak in the community reaches 0.95 ($t_{95}$) for a set of scenarios corresponding to potential Delta variant parameter combinations. The vaccine coverage values correspond to the proportion of individuals in the community (outside of quarantine) who are vaccinated. Higher vaccine coverage and efficacy reduces the probability of an outbreak occurring, given a quarantine breach event. Possible threshold behaviour is observed for high coverage and efficacy, with vaccine efficacy of 90\% producing $t_{95}$ values that increase by more than one order of magnitude as coverage rises from 0.1 to 1.0 (full vaccine coverage). The baseline scenario (black line) shows $t_{95}$ with no vaccination ($VE = 0$) and $R_0 = 3$.}
    \label{fig:t_95}
\end{figure}

We present $t_{95}$ values as a function of vaccine coverage (Figure \ref{fig:t_95}), which demonstrates the effect of mass vaccination on outbreak risk, for each combination of $VE$ and $R_0$. The results illustrate that vaccine efficacy is a crucial determinant of outbreak risk. Even for high coverage, a vaccine with efficacy below 60\% is not sufficient to dramatically reduce time required for outbreaks to occur. On the other hand, a possible threshold behaviour with coverage is observed for high vaccine efficacy (80\% or higher), with $t_{95}$ values increasing by an order of magnitude for high coverage levels. This threshold effect appears to persist even for very high transmission rates (Figure \ref{fig:t_95}f, $R_0 = 10$). The sensitivity of outbreak risk to $VE$ emphasise the importance of accurately estimating this crucial parameter (vaccine efficacy against infection and onward transmission).

\subsection{Sensitivity analysis: incubation period}
Our sensitivity analysis of the incubation period demonstrated that shorter incubation periods increase the effectiveness of the quarantine system because (i) test sensitivity increases more rapidly, and (ii) individuals are infectious for shorter periods of time. These results emphasise that shorter incubation periods can make it easier to detect infections in closed systems like quarantine facilities. This reduces the risk to the community for the same quarantine length-of-stay. 

\subsection{Sensitivity analysis: Vaccine efficacy against transmission from breakthrough infections}
The results demonstrated in Figure \ref{fig:net_FoI_vs_R0_x_VE} assume no effect of vaccination on the capacity for vaccinated individuals who become infected to transmit the virus ($V_T = 0,~V_I = VE$). The alternate assumption that efficacy against onward transmission is equivalent to the total $VE$ represents a plausible upper bound, that we investigated in a sensitivity analysis ($V_T = VE,~V_I = 0$). The results in Figure \ref{fig:scan_R0xVE_VEimax} demonstrate that under this alternate (optimistic) assumption, the vaccine efficacy required to maintain baseline risk levels falls by about 20\%. For example, an $R_0$ of 6 would require a 40\% effective vaccine in order to maintain baseline outbreak risk while an $R_0$ of 8 would require a 50\% effective vaccine. 

\section{Discussion}

The vaccines that have been developed against SARS-CoV-2 remain highly effective at preventing severe disease. However, their efficacy against infection has decreased against the Delta variant of the virus, and efficacy may deteriorate further with the continued emergence of new variants \cite{Puranik2021comparison,pouwels2021impact}. In the context of border quarantine, the capacity to limit transmission is the key consideration when determining how best to manage new arrivals, some of whom may be infected (asymptomatic or pre-symptomatic). This is because the primary purpose of a border quarantine system is to prevent infectious individuals from entering the community. The management of clinical cases within the quarantine system is facilitated by regular surveillance, efficient case detection, and the allocation of medical resources. Therefore, the utility of vaccination within the context of a quarantine system is not equivalent to the utility of mass vaccination in the context of a large outbreak. In large outbreaks, efficacy against clinical severity reduces hospital case loads and deaths, mitigating the public health burden and human cost, even if transmission continues. However, in modern quarantine systems, the operational goal is to identify and isolate cases to limit transmission and keep the required duration of quarantine to a minimum. 

The 14-day minimum stay commonly practiced during the COVID-19 pandemic was implemented due to the long incubation period of the disease. Within 14 days, a case arriving while pre-symptomatic would be likely to display symptoms. Testing of arrivals through RT-PCR accelerates the process of case detection and facilitates earlier management. When these case detection efforts are unsuccessful, transmission within the quarantine environment at best leads to extended stay conditions for cases and their contacts. At worst, it leads to the discharge of pre-symptomatic infectious individuals who receive their exit test early in their infection and do not yet have high enough viral load for case confirmation. Therefore, the primary benefit of vaccination in the context of quarantine facilities is in limiting transmission. 

In this context, the increased transmissibility and decreased vaccine efficacy against transmission associated with the Delta variant requires a re-evaluation of risk. Our results demonstrate that vaccination may allow quarantine systems to remain effective. On the other hand, quarantine requirements for vaccinated travellers must remain stringent due to the increased transmissibility of the virus. To emphasise the implications of this result, had conditions remained consistent with the Alpha variant ($R_0 \approx 3$, $VE\approx 90\%$), Figure \ref{fig:net_FoI_vs_R0_x_VE} indicates that vaccination would have decreased border quarantine breach risk to 3\% of baseline. In effect, the vaccinated quarantine pathway {\it would have} allowed the number of arrivals to increase by a factor of approx. 30 (assuming sufficient system capacity), while maintaining baseline community exposure levels.  

Under the existing circumstances, our analysis suggests that quarantine policies for vaccinated individuals will need to approximate those that were used for unvaccinated cohorts prior to the emergence of the Delta variant. Unvaccinated cohorts of travellers, on the other hand, will pose a much greater risk of quarantine breach events than they did previously (e.g., increased by a factor of 3 for $R_0 = 6$, Figure \ref{fig:net_FoI_vs_R0_x_VE}). 

Ultimately, the level of stringency in requirements for quarantine of new arrivals should be assessed as a function of the prevalence of viral variants between jurisdictions. Where these relative levels are similar for variants of concern, there is little justification for limitations on travel. However, with the spatially-localised emergence of new variants, quarantine systems must be capable of rapidly responding to slow, or eliminate, the global diffusion of variants. This is particularly true for variants with increased transmissibility and clinical severity (of which the Delta variant of SARS-CoV-2 is a primary example).

\section{conclusion}

To summarise, our results demonstrate that, in the context of SARS-CoV-2, border quarantine systems cannot be used to compensate for low levels of community vaccination. For the Delta variant, this is true even when all individuals within the quarantine environment are vaccinated. This is because the Delta variant is more transmissible and likely to produce breakthrough infections in individuals vaccinated against the ancestral lineage. Our findings illustrate a key aspect of the drawn-out global battle to mitigate the public health crisis produced by COVID-19. Just as regions with low community prevalence were becoming confident that international travel could increase for vaccinated individuals, the virus changed to become more transmissible, and to partially avoid vaccine-induced immunity. Our results show that these changes nullified the prospective benefits of quarantine systems in terms of international travel volumes. However, it is important to emphasize the alternative scenario: had a vaccine not become available prior to the emergence of the Delta variant, the capacity for existing border quarantine systems to mitigate outbreaks would have dramatically deteriorated. Currently, comprehensive vaccination in quarantine facilities is allowing countries such as Australia, New Zealand, and China to continue allowing low levels of international travel. 

Moving forward, the expansion of quarantine systems should focus on preparing for future variants of COVID-19, and ultimately, for future pandemics with higher clinical severity and infection fatality ratios. COVID-19 has revealed the unprecedented capacity for populations around the world to dramatically alter their behaviour to prevent disease spread. Quarantine systems amplify the payoff of these population-wide responses by limiting incursions. The coupling of border quarantine measures with elimination strategies buys critical time for the development of vaccines and effective clinical practices.

\section{Acknowledgements}
{\bf Funding:} This work was directly funded by the Australian Government Department of Health Office of Health Protection. Additional support was provided by the National Health and Medical Research Council of Australia through its Centres of Research Excellence (SPECTRUM, GNT1170960) and Investigator Grant Schemes (JMcV Principal Research Fellowship, GNT 1117140). MJL is supported by an NHMRC Project Grant (GNT1156742) {\bf Conflicts of interest:} The Authors declare no conflicts of interest. {\bf Author Contributions:} CZ, NG, and J. McVernon designed the model of quarantine facility environments; CZ, FS, DJP, and J. McCaw designed the individual-based model of COVID-19 disease progression and test sensitivity; MJL, DJP and FS designed the outbreak branching process model; CZ implemented the individual-based models and analysed outputs; ML implemented the outbreak model and analysed outputs; All authors contributed to manuscript composition and study design.

\section{Source code and data availability}
Source code for the individual-based models of COVID-19 and the raw data used to produce Figure \ref{fig:net_FoI_vs_R0_x_VE} is available at: \url{github.com/cjzachreson/COVID-19_Quarantine_ABM}.\\
Source code for the branching process model is available at: \url{github.com/MikeLydeamore/COVIDQuarantine/tree/quarantinemodel}.\\
All data presented in this work can be reproduced from the source code above. It can also be made available upon request to the corresponding author.

\bibliography{references.bib}

\newpage

\newcommand{\beginsupplement}{%

 \setcounter{table}{0}
   \renewcommand{\thetable}{S\arabic{table}}%
   
     \setcounter{figure}{0}
      \renewcommand{\thefigure}{S\arabic{figure}}%
      
      \setcounter{page}{1}
      \renewcommand{\thepage}{S\arabic{page}} 
      
      \setcounter{section}{0}
      \renewcommand{\thesection}{S\arabic{section}}
      
      \setcounter{equation}{0}
      \renewcommand{\theequation}{S\arabic{equation}}
     }


\beginsupplement

\FloatBarrier
{\bf \Large{Supporting Information}}

\section{Model details}

This work utilised two distinct models of COVID-19 disease transmission. One of these is an abstract model, implemented as a branching process to approximate community transmission dynamics. This is applied to quantify the tendency for quarantine breach events to initiate outbreaks, and to characterise those outbreaks. The other model is an agent-based simulation describing disease progression and transmission on the individual level. This is implemented in two scenarios, one used for calibration of fundamental parameters, and the other for investigation of quarantine system efficacy. The calibration scenario is homogeneous, and allows characterisation of the basic reproductive ratio, over-dispersed secondary case rates, and timing of transmission detection events in an effectively open system. The quarantine scenario is designed to reproduce the general features of controlled border screening environments, and implements a defined population structure.

\subsection{Outbreak transmission (Branching process) model}
To simulate infectious individuals entering the community and potentially seeding an outbreak, we use a branching process community transmission model with an inhomogenous offspring distribution. Branching process models are considered in terms of generations, and each generation acts independently from the last. We term `generation zero' to be the index case that leaves quarantine.

\subsubsection{Transmission potential of cases in the community}
We define the \emph{transmission potential}, which represents the expected number of cases caused by a single case in the community. The transmission potential combines the biological features of the virus, vaccine status of the index case, vaccination coverage in the wider community and vaccine efficacy parameters.

We break down vaccine efficacy into two components: efficacy against infection, denoted $V_I$, and efficacy against onward transmission, denoted $V_T$. The overall reduction in transmission is given by, 
$$V=(1-V_I) (1-V_T).$$
For a given starting transmission potential, $TP_0$, the effective transmission potential in the population, $TP_p$, is given by,
$$TP_p=TP_0 \cdot (1-cV),$$ where $c$ is the proportion of the population vaccinated.
For a breach caused by a vaccinated traveller, the first generation transmission potential is given by, $$TP_u=TP_0 \cdot (1-cV_I).$$
For a breach caused by an unvaccinated traveller, the first generation transmission potential is given by, $$TP_v=TP_0 \cdot ( 1-c V_I)(1-V_T).$$

\subsubsection{Secondary case distribution and public health measures}
The number of secondary infections generated by a single case is taken from a negative binomial distribution, with $n=0.1$ samples, and the probability of success being $p=0.1/(0.1 + TP)$, in order to match the dispersion in the number of secondary cases (fixed at $0.1$), and the expected number of secondary cases.

The calendar times at which these infections occur is drawn from a Weibull distribution with mean $5.5~\text{days}$ and standard deviation $1.8$, in line with literature estimates \cite{ferretti2020quantifying}.

If the case is the index case and thus in quarantine, only infections that occur \textit{after} the individual has left quarantine are retained. If the individual enters isolation at some point, as per worker protocols, then only infections that occur \textit{before} the individual has entered isolation are retained. All future generations are assumed to not be in quarantine or have any proactive isolation applied to them.

\subsubsection{Probability of a breach becoming an outbreak}
To calculate the probability of a single breach event becoming an outbreak, we simulate $10^5$ breaches into the community. The index cases for these breaches are chosen randomly from the quarantine model. The probability is calculated separately for travellers and workers.

A breach is defined as an outbreak if it reaches at least five cumulative cases before extinction.

\subsubsection{Time until significant breach event}
The time between breach events are assumed to occur at a frequency defined by the quarantine model. To scale this time according to the number of travellers and the prior probability of arriving to the system infected, we construct a scaling factor, $$s=\frac{A}{\bar{A}} \frac{I}{\bar{I}},$$ where $A$ is the targeted number of arrival, $I$ is the targeted prevalence of the arrival's country, $\bar{A}$ is the number of arrivals modelled in the quarantine model and $\bar{I}$ is the prevalence of infection assumed in the quarantine model. The time between breaches obtained from the quarantine model is divided by $s$.

A breach is determined to be an outbreak according to a binomial distribution, with probability of success defined by the probability of causing an outbreak. Traveller and worker breach events are chosen in the same proportion at which they occur from the quarantine model.

\subsection{Individual-based COVID-19 disease model}
For the agent-based model of quarantine scenarios, we developed a model of COVID-19 disease and transmission designed to match three salient features: 
\begin{enumerate}
    \item The distribution of delays between symptom onset of a primary case and transmission to secondary cases. Following the definition used by Ferretti et al., we refer to this quantity as Time from Onset of Symptoms to Transmission (TOST) \cite{ferretti2020timing}.
    \item The household secondary attack rate, and secondary case dispersion.
    \item The dependence of RT-PCR test sensitivity on time from symptom onset. 
\end{enumerate}
Matching these distributions with an individual-based model required the definition a detailed model of disease natural history on the within-host level. The distribution of incubation periods is a key ensemble statistic that informs the model dynamics (see below). There are many possible implementations of within-host models that could generate the required ensemble statistics. Our specific choices follow the logic that virus initially grows exponentially, until recognition by the host immune systems triggers the onset of symptoms (the end of the incubation period). This immune response produces an exponential decline in viral load until recovery occurs, at which time no replication-competent viral shedding is possible. The sections below detail the specifics of the agent-based model. 


\subsubsection{The basic reproductive number $R_0$}\label{supsec:R0}

\begin{figure}
    \centering
    \includegraphics[width = \textwidth]{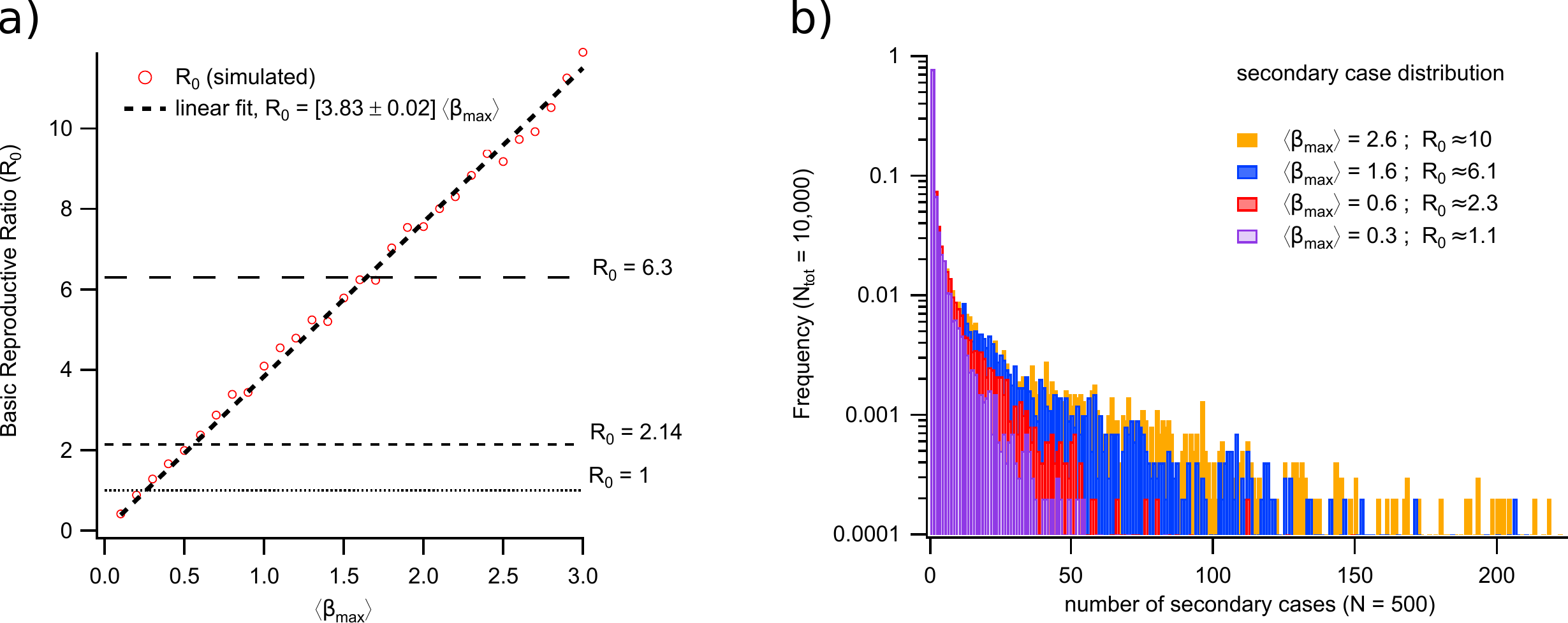}
    \caption{Calibration of the basic reproductive number $R_0$. (a) Linear dependence between the global transmission scalar $\langle \beta_{max}\rangle$ and the derived parameter $R_0$. (b) The distribution of secondary cases produced for different values of $R_0$, these are effectively drawn from a negative binomial distribution with dispersion parameter (``number of successes" $r = 0.1$), and mean $R_0$.}
    \label{fig:R0_vs_beta_N500}
\end{figure}

The transmission rate scalar for each individual $\beta_{max} \sim \text{Gamma}(\kappa, \theta)$ controls the transmission rate for any given transmission environment. The shape parameter $\kappa = 0.1$ translates directly to the dispersion parameter of the derived secondary case distribution (distributed as a negative binomial, see below). The scale parameter $\theta = \langle \beta_{max}\rangle / \kappa$ is a function of the mean peak force of infection $\langle \beta_{max}\rangle$, which is proportional to $R_0$ in the calibration model (Figure \ref{fig:R0_vs_beta_N500}). 

To calibrate the basic reproductive number $R_0$, we performed a systematic scan of $\langle \beta_{max}\rangle$, keeping all other parameters constant. To produce a generic calibration of the basic reproductive number $R_0$, we performed this scan on an unstructured population with N = 500 individuals, and simulated a large ensemble of single transmission generations without interaction effects (N = 10,000 instances). For each instance, the index case properties are sampled from the parameter distributions specified and transmission is simulated until recovery of the index case. We then count the secondary cases produced, ignoring the additional force of infection produced by secondary cases. The average of these values approximates $R_0$, and we observe a linear relationship with the control parameter $\langle \beta_{max}\rangle$ (Figure \ref{fig:R0_vs_beta_N500}). To set the value of $R_0$ in a given scenario, we use the line of best fit $R_0 = 3.83 \langle \beta_{max}\rangle$ to determine the corresponding value of $\langle \beta_{max}\rangle$ required to produce the desired value of $R_0$. 

\subsubsection{Secondary case dispersion} 

\begin{figure}
    \centering
    \includegraphics[width = 0.6\textwidth]{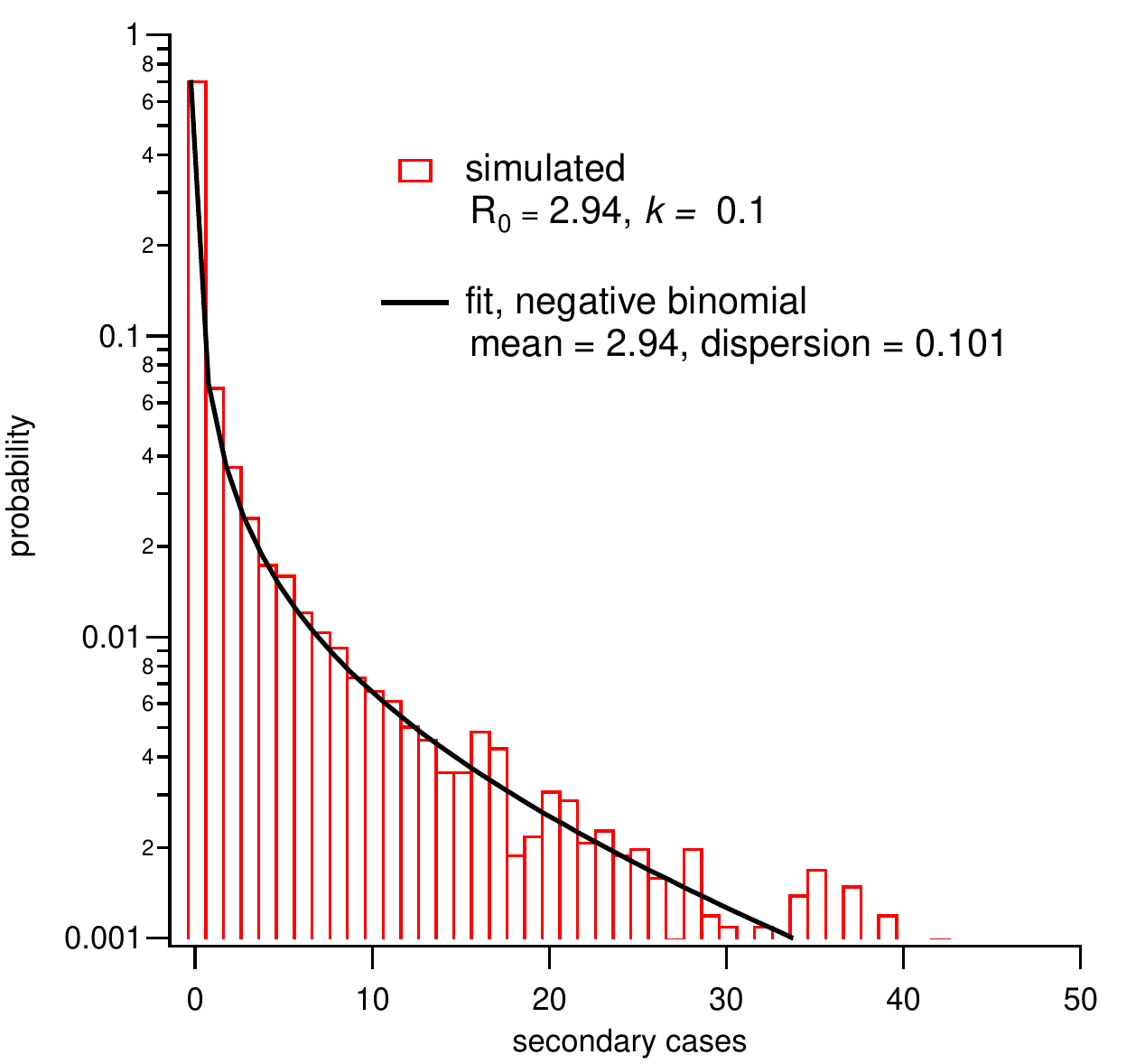}
    \caption{The secondary case distribution of the calibration model follows a negative binomial with mean $R_0$. The dispersion parameter ($r$) is controlled by the shape parameter ($\kappa$) of the Gamma distribution from which maximum force of infection ($\beta_{max}$) is sampled for each individual. }
    \label{fig:nbinom}
\end{figure}

The distribution of secondary cases produced by the index case ensemble used to compute $R_0$ conforms to a negative binomial distribution with dispersion parameter $r = 0.1$ (Figure \ref{fig:nbinom}). This is implemented as follows: for each index case, independent transmission probabilities produce Poisson-distributed secondary case numbers. The transmissibility parameter for each index case $\beta_{max}$ is sampled from a Gamma distribution, so the secondary case numbers aggregated over all index cases are effectively drawn from an ensemble of Poisson distributions with Gamma-distributed rate parameters. This gives a negative binomial with dispersion $r = \kappa$ \cite{lloyd2005superspreading}. Therefore, the model controls secondary case dispersion directly by assigning $\beta_{max}$ a Gamma-distributed random variable. The choice of dispersion does not alter the calibration of $R_0$, which describes only the average number of secondary cases. 

\subsubsection{Household secondary attack rate}
\begin{figure}
    \centering
    \includegraphics[width = \textwidth]{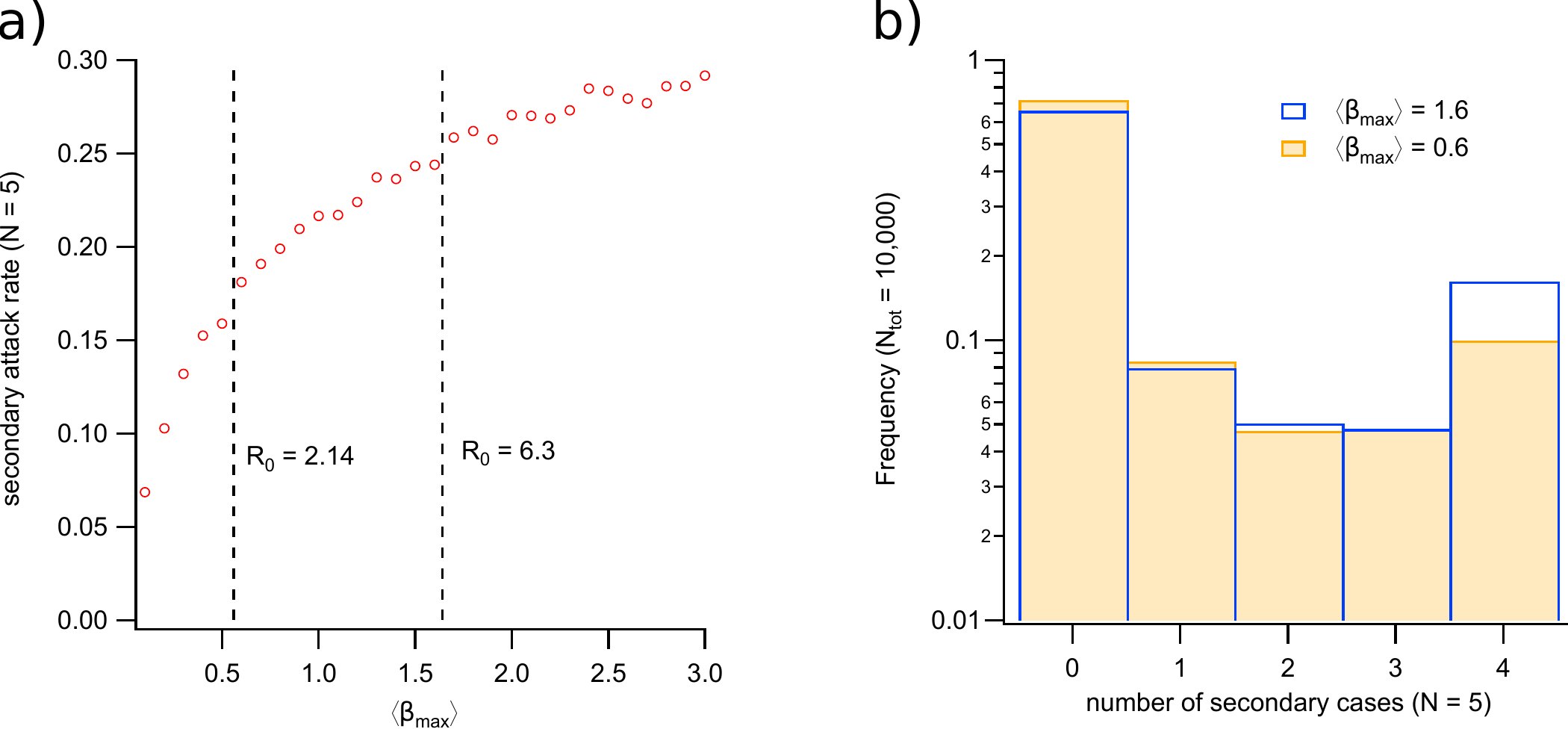}
    \caption{Secondary attack rate values produced by the calibration model for transmission within small groups (N = 5). (a) Secondary attack rate as a function of the global transmission scalar $\langle \beta_{max}\rangle$. The values plotted correspond to the number of secondary infections produced on average divided by the number of contacts (N - 1 = 4). (b) Frequency distributions of the secondary case numbers produced by 10,000 independent trials each corresponding to a single index case over a single generation of transmission to four close contacts. The values of $\langle \beta_{max}\rangle$ chosen for (b) correspond broadly to estimates of $R_0$ for the ancestral lineage of SARS-CoV-2 ($\langle \beta_{max}\rangle$ = 0.6, $R_0 = 2.14$) and the Delta variant ($\langle \beta_{max}\rangle$ = 1.6, $R_0 = 6.3$).   }
    \label{fig:SAR_vs_beta_N5}
\end{figure}

To check that our parameterisation of $R_0$ and $r$ produce reasonable correspondence with observed household secondary attack rates, we ran the calibration model setting the population size N = 5. Under these conditions, the number of possible transmissions is akin to a generic number of household contacts (N - 1) and individuals with high transmissibility have restricted transmission potential. This calibration produces average household secondary attack rates in a range between 10\% and 25\%, consistent with observations of COVID-19 transmission among household contacts (Figure \ref{fig:SAR_vs_beta_N5}) \cite{madewell2020household,kang2021transmission}. The low end of the range (10\%) is typical for the ancestral lineage of SARS-CoV-2, and this increases consistently with $R_0$, reaching 25\% for $R_0 = 6.3$, corresponding to the Delta variant.  

\FloatBarrier

\subsubsection{Individual-based disease transmission model}
\begin{figure}
    \centering
    \includegraphics[width = 0.65\textwidth]{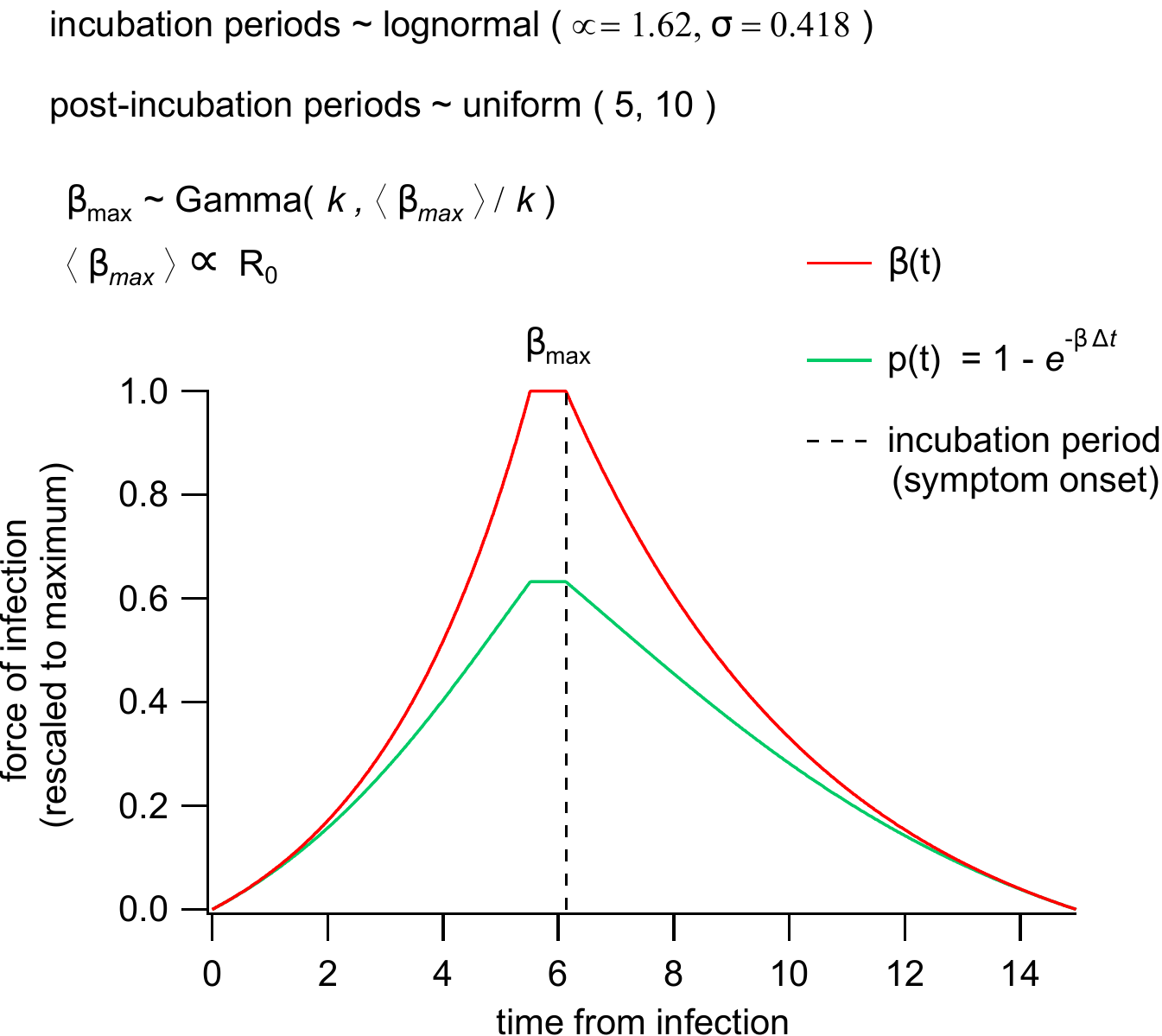}
    \caption{An example of the force of infection produced by an infected individual as a function of time from infection. The force of infection increases exponentially from the time of infection to the start of the plateau, at which point it reaches its maximum value. The plateau phase lasts until the end of the incubation period and has a duration equal to 1/10 of the total incubation period. Incubation periods are drawn from a log-normal distribution. During the post-incubation period, infectiousness decreases exponentially until it reaches a cutoff at the time of recovery. The duration between the end of the plateau phase and recovery is drawn from a uniform distribution bounded between 5 and 10 days. The y-axis values are re-scaled to $\beta_{max}$, which is drawn from a Gamma distribution parameterised for a specified $R_0$ value as described in Section \ref{supsec:R0}.}
\label{fig:beta_vs_t}
\end{figure}

Our individual-based model of within-host viral dynamics follows a three-part function, the precise form of which is determined by each individual's incubation period. An individual trajectory of infectiousness as a function of time from exposure $\beta(t)$ is illustrated in Figure \ref{fig:beta_vs_t}. This function features an initial exponential increase, followed by a brief plateau and subsequent exponential decline. The start of the exponential decline phase corresponds to the end of individual's incubation period (onset of symptoms). Note that individuals become infectious immediately after they are infected, with no latent period preceding the onset of infectiousness. In our model, we assume that 1/3 of infected individuals never develop symptoms, which reduces the probability they will be detected within the quarantine system. In our model we use the same functional form of $\beta(t)$ to describe all infections, regardless of whether or not they are asymptomatic. It may be that asymptomatic individuals are marginally less contagious, in which case our model would overestimate the number of secondary cases they produce \cite{johansson2021sars}. In this work, we opted for the conservative assumption that no difference exists.

The within-host dynamics are implemented as a piecewise function, with a plateau between growth and decay phases describing $\beta(t)$, the time-dependent infectiousness of an individual: 
\begin{equation}\label{eq:infectivity}
\beta(t)  =
 \begin{cases} 
     \frac{\beta_{\text{max}}}{V_{\text{max}}} \big[\exp(k_{1}t) - 1\big] & t\leq t_{\text{inc}} - T_{p} \\
      \beta_{\text{max}} & t_{\text{inc}} - T_p \leq t < t_{\text{inc}}\\
      \beta_{\text{max}}\big[\exp\big(k_{2}[t-t_{\text{inc}}]\big) - [V_{\text{max}}]^{-1}\big] & t \geq t_{\text{inc}}
   \end{cases}\,,
\end{equation}
where $t_{\text{inc}}$ is the incubation period of the individual, $T_p$ is the duration of the infectiousness plateau (set equal to $0.1t_{\text{inc}}$), and $\beta_{\text{max}}$ is the maximum infectiousness of that individual. The parameter $V_{\text{max}}$ is a scaling factor controlling the shape of the growth curve (smaller values of $V_{\text{max}}$ produce broader growth and decay functions, higher values produce steeper growth and decay). The rate parameters $k_1$ and $k_2$ are determined by the value of $V_{\text{max}}$, and the duration of incubation and post-incubation periods: 
\begin{equation}
    k_1 = \ln(V_{\text{max}})~[t_{\text{inc}} - T_p]^{-1}\,,
\end{equation}
and
\begin{equation}\label{eq:k2}
    k_2 = \ln\big([V_{\text{max}}]^{-1}\big)~t_{\text{r}}^{-1}\,,
\end{equation}
where $t_r$ is the time between symptom onset and recovery, which is drawn uniformly at random from the range [5d, 10d], to approximately match the duration of replication-competent viral shedding after symptom onset \cite{wolfel2020virological}. For an interaction between infected individual $i$ and susceptible individual $j$, the probability of transmission is computed as:
\begin{equation}
    p_{ij}(t) = 1 - \exp(-\beta(t, i) \sigma_{ij} \Delta t)\,,
\end{equation}
where $\beta(t, i)$ is the force of infection produced by an infected individual $i$ at time $t$ since infection, $\Delta t = 0.1~\text{days}$ is the duration of a discrete time step, and $\sigma_{ij}$ is a scaling factor that incorporates the effects of contact frequency and intensity for a given transmission event between infected individual $i$ and susceptible individual $j$:
\begin{equation}
    \sigma_{ij} = F_{ij} h_j^{-1}\,,
\end{equation}
where $F_{ij}$ is a context-specific transmission mitigation factor, and $h_j$ is the number of contacts of a given type in the local mixing environment. In the calibration model, $F_{ij} = 1$ and $h_j = N - 1$. In the full quarantine model, $F_{ij}$ is given as follows:
\begin{itemize}
    \item For close contacts between travellers in the same group, \\$F_{ij} = 1$, and $h_j = n_i - 1$, where $n_i$ is the number of close contacts in the same group as individual $i$ ($n = 4$ initially but can decrease if members are moved into isolation).
    \item For interactions between travellers in different close contact groups,\\ $F_{ij} = 0.01$ and $h_j = \sum\limits_{g | i \notin g}{n_g}$ is the number of travellers in the quarantine system who are not isolated and who are not in the same group $g$ as individual $i$).
    \item For interactions between infected travellers and susceptible workers, \\$F_{ij} = 0.01$ and $h_j = n_w$, where $n_w$ is the number of workers {\it present} in the facility at time $t$. 
    \item For interactions between infected workers and susceptible travellers, \\$F_{ij} = 0.01$, $h_j = n_{tot}$ where $n_{tot}$ is the number of travellers not in isolation.
    \item For interactions between infected workers and susceptible workers, \\$F_{ij} = 0.1$ and $h_j = n_{w} - 1$, where $n_w$ is the number of workers {\it present} in the facility at time $t$.  
\end{itemize}


\FloatBarrier

\subsubsection{Time from Onset of Symptoms to Transmission (TOST)}
\begin{figure}
    \centering
    \includegraphics[width = 0.65\textwidth]{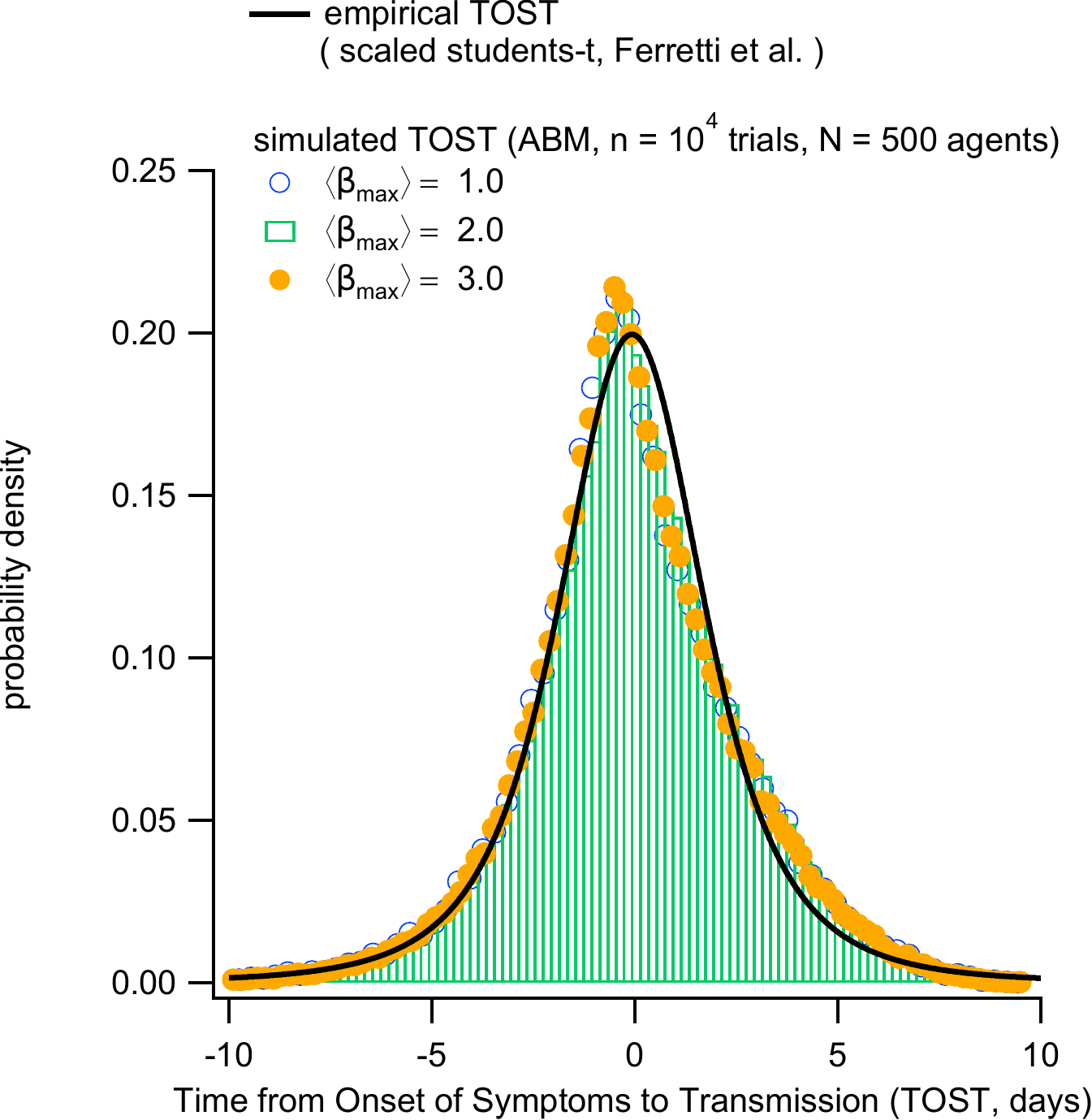}
    \caption{Time from Onset of Symptoms to Transmission (TOST) distributions produced by the calibration model with N = 500 individuals in a fully-connected (homogeneous) contact network. The TOST distribution does not depend on the transmissibility of the disease, and matches closely with the best-fit model from Ferretti et al. \cite{ferretti2020timing}.}
    \label{fig:tost}
\end{figure}

The functional form used for $\beta(t)$ was developed to match the ensemble distributions of time from onset of symptoms to transmission (TOST), reported by Ferretti et al. \cite{ferretti2020timing}. A comparison of the statistics produced by the calibration model and the distribution reported by Ferretti et al. is shown in Figure \ref{fig:tost}. The qualitative match between our model's case statistics and the empirical TOST distribution is sensitive to the choice of individual disease trajectory function (Equations \ref{eq:infectivity} through \ref{eq:k2}).  

\FloatBarrier

\subsubsection{Time from Onset of Symptoms to Detection (TOSD)}

The model of test sensitivity discussed in the following subsections is designed to capture the variability in RT-PCR test sensitivity as a function of time relative to symptom onset. These models are based on a study by Hellewell et al. \cite{hellewell2021estimating}. Here, the general objective is to describe test sensitivity as a piecewise sigmoid with a single breakpoint, $T_c$. Before the breakpoint sensitivity increases ($t < T_c$, where $t$ represents time from infection). After the breakpoint ($t > T_c$), sensitivity decreases slowly. In this case, the function is expressed as a logistic regression: 

\begin{equation}
p (\text{positive} | \text{infected}) =
 \begin{cases} 
      [1 + \exp(-(b_1 + b_2\tau))]^{-1} & t\leq T_c \\
      [1 + \exp(-(b_1 + b_2\tau + b_2b_3\tau)]^{-1} & t > T_c
   \end{cases}\,,
\end{equation}
where $\tau = t - T_c$, the maximum test sensitivity depends only on $b_1$ as $[1 + \exp(-b_1)]^{-1}$, the initial growth of sensitivity is controlled by $b_2$, and the decay of sensitivity after $T_c$ is given by $b_2$ and $b_3$. The parameter ranges (95\% credible intervals) provided in Hellewell et al. are shown in table \ref{table:hellewell_params}.

\bgroup
\def\arraystretch{1.25}
\begin{table}
    \centering
    \vspace{0.5cm}
    \resizebox{0.5\textwidth}{!}{%
    \begin{tabular}{| c | c |}
    \hline
         ~parameter~ & ~95\% CI~ \\
       \hline
       $T_c$ & [2.01, 5.11] \\
       \hline
        $b_1$ &  [0.8, 2.31]  \\
      \hline
        $b_2$ & [1.26, 3.47]\\
      \hline
        $b_3$ & [-1.2, -1.05] \\
      \hline
    \end{tabular}
    }
    \caption{Parameter ranges used in Hellewell et al., to describe the progression of test sensitivity during COVID-19 infections in healthcare workers. In the model reported here, these parameter ranges were adjusted to account for individual variability in incubation periods.  }
    \label{table:hellewell_params}
\end{table}
\egroup

The parameter ranges shown in Table \ref{table:hellewell_params} were produced by fitting to ensemble data. That is, the study examined a timeseries of tests and test results for a cohort of 27 individuals, and aggregated this timeseries in order to fit the parameters of the model. Therefore, the parameter ranges above are not suitable for all individual test sensitivity trajectories. Our individual-based model translates the specified functional form (piece-wise logistic function) and parameter ranges into a set of trajectories describing test sensitivity as a function of the incubation period (time between infection and symptom onset) for each individual in the system. By doing so, we reconcile the ensemble characteristics observed by Hellewell et al., with the requirements of individual trajectories. In our model, the requirements are specified as follows: 
\begin{itemize}
    \item {An individual may not test positive before they are infected.}
    \item {The peak of test sensitivity must have some correspondence to the timing of the peak viral load (peak infectiousness).}
    \item{The probability of detection should be high during the first week of symptoms.}
\end{itemize}
These requirements have the following implications for the model framework specified by Hellewell et al.: \begin{itemize}
    \item{the specified range of $T_c$ will not fit trajectories with longer-than-average incubation periods (average incubation period is 5.5 days).}
    \item{The rate of increase ($b_2$) must be higher for individuals with shorter incubation periods (this imposes a correlation when specifying individual trajectories).}
\end{itemize}

To address these discrepancies, we relax the parameter restrictions in Table \ref{table:hellewell_params}, and set the breakpoint position relative to an individual's incubation period. The delay between the breakpoint $T_c$ and the onset of symptoms $t_{inc}$ is distributed in the range $[0, 4.11]$.
The rationale for this choice is based on the observation that the upper 95\% credible interval reported by Hellewell, $T_c = 5.11$, is approximately equivalent to the median incubation period. Here, we infer that this corresponds to using the incubation period as an upper limit for $T_c$. Additionally, our model correlates the length of the delay with the incubation period of each agent by quantile matching, to ensure the that $T_c$ is always positive:
\begin{equation}
T_{c,i} = t_{inc,~i} - 4.11q_i\,,
\end{equation}
where $i$ denotes a specific individual, $q_i$ is the value of the incubation period CDF evaluated at $t_{inc}(i)$, and 4.11 is the range of possible delays (in units of days) between peak test sensitivity and symptom onset.

\begin{figure}
    \centering
    \includegraphics[width = 0.5\textwidth]{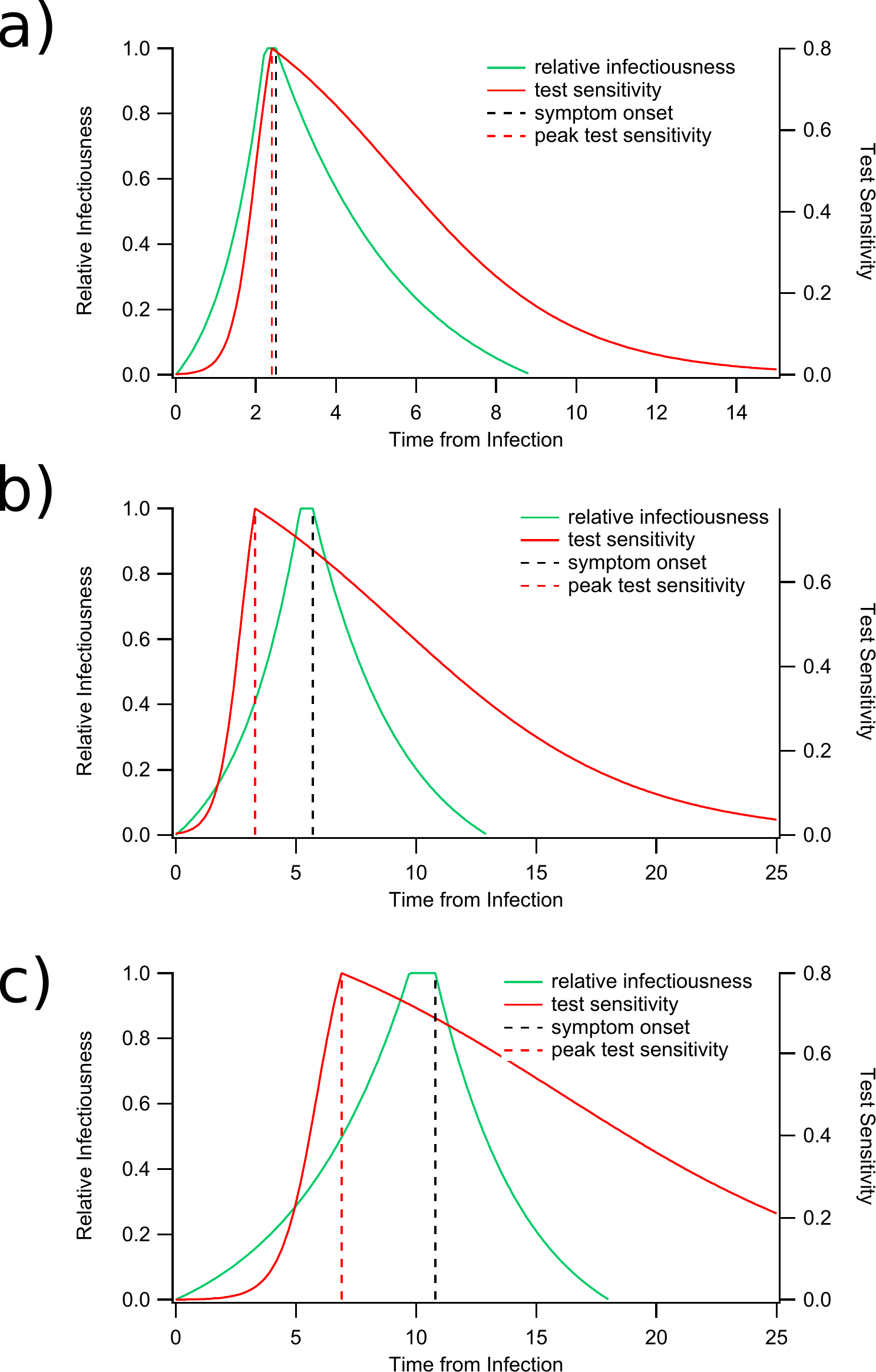}
    \caption{Individual trajectories of test sensitivity and force of infection as a function of time since exposure. (a) Short incubation periods produce alignment of peak test sensitivity and peak infectiousness, both of which occur close to the onset of symptoms. (b) Average incubation periods produce a test sensitivity peak that lags symptom onset and peak infectiousness by several days, and extend the tail of the test sensitivity curve after symptom onset. (c) Long incubation periods correspond to longer lag times between peak test sensitivity and symptom onset, and further extend the tail of the test sensitivity trajectory.   }
    \label{fig:tsens_i_traj}
\end{figure}

\begin{figure}
    \centering
    \includegraphics[width = \textwidth]{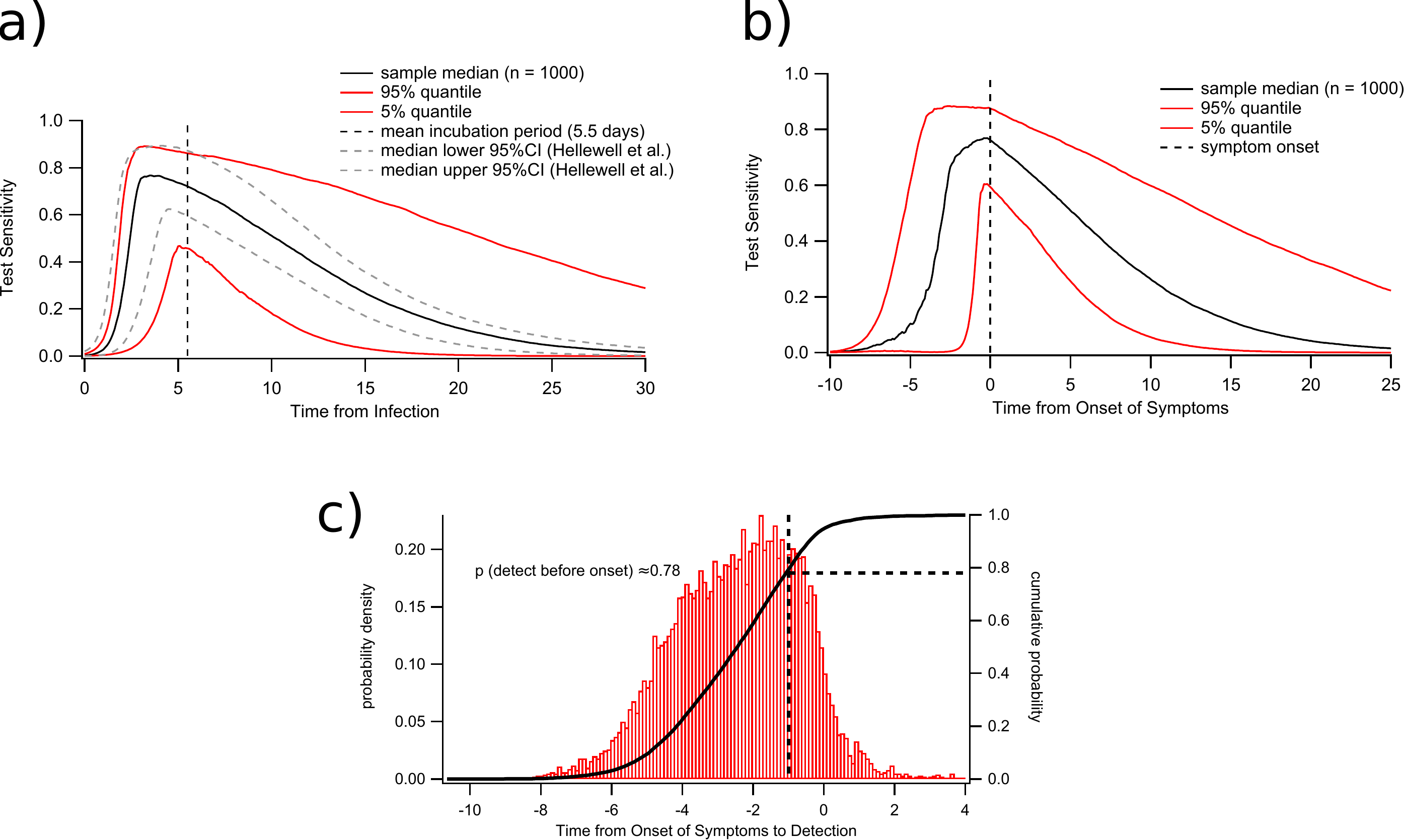}
    \caption{Ensemble statistics of test sensitivity as a function of time from infection, aggregated from the Agent based model. (a) Test sensitivity as a function of time from infection. Grey dashed lines give the 95\% credible intervals of the mean from Hellewell et al., while the solid black trace shows the ensemble average produced by 1000 samples from the Agent-based model. Solid red traces give the 5\% and 95\% quantiles from the Agent-based model. (b) Test sensitivity as a function of time from symptom onset produced by the Agent-based model. The black trace gives the ensemble average while the red traces give 5\% and 95\% quantiles. The vertical dashed lines in (a) and (b) represent the onset of symptoms. (c) Time from Onset of Symptoms to Detection given daily tests via RT-PCR, derived from the Agent-based model. The solid trace indicates the cumulative distribution, while the dashed lines indicate the proportion of cases (0.78) detected more than one day prior to symptom onset (assuming a test turnaround of 1 day, this corresponds to the proportion detected prior to symptom onset).  }
    \label{fig:tsens_ensemble}
\end{figure}

To provide a better match to ensemble statistics, we also chose to impose a negative correlation between individual incubation periods and the term ($b_2$) that specifies the growth rate of test sensitivity before the peak, and (along with $b_3$) the decay rate of test sensitivity after the peak:
\begin{equation}
b_{2,i} = 1.26 - 2.21(1 - q_i) \,,
\end{equation}
where 1.26 is the lower bound for $b_2$ and 2.21 is the range of $b_2$ values given in Table \ref{table:modified_params_3p5p2}. The modified parameter ranges are shown in Table \ref{table:modified_params_3p5p2}, samples of individual trajectories are shown in Figure \ref{fig:tsens_i_traj}, and a comparison of ensemble statistics is shown in Figure \ref{fig:tsens_ensemble}. On the other hand, the $b_1$ and $b_3$ parameters are chosen uniformly at random for each individual from the ranges specified in $\ref{table:modified_params_3p5p2}$.



\bgroup
\def\arraystretch{1.25}
\begin{table}
    \centering
    \vspace{0.5cm}
    \resizebox{0.5\textwidth}{!}{%
    \begin{tabular}{| c | c |}
    \hline
         ~parameter~ & ~95\% CI~ \\
       \hline
       $T_c$ & $t_{\text{inc}}$ - [0, 4.11] \\
       \hline
        $b_1$ &  [0.8, 2.31]  \\
      \hline
        $b_2$ & [1.26, 3.47]\\
      \hline
        $b_3$ & [-1.14, -1.05] \\
      \hline
    \end{tabular}
    }
    \caption{Parameter ranges used in order to fit the ABM ensemble statistics to those reported by Hellewell et al.. }
    \label{table:modified_params_3p5p2}
\end{table}
\egroup

The model as-implemented approximately matches the proportion of infections detected before symptom onset, given daily RT-PCR test, estimated by Hellewell et al. (approx. 75\%). The corresponding distribution of time from onset of symptoms to detection (TOSD), is shown in Figure \ref{fig:tsens_ensemble}(c). In this system, daily testing detects approximately 78\% of cases prior to symptom onset. 

\FloatBarrier

\subsection{Model of quarantine environment} 

The model of quarantine system in which transmission occurs, explicitly represents individual travellers and quarantine workers. We describe the model in terms of an Input layer, Filter layer and Output layer.

\subsubsection{Input layer}
Arriving travellers are characterised in terms of:
\begin{itemize}
    \item the proportion of arriving travellers who are infected
    \item the proportion of arriving travellers who are vaccinated
\end{itemize}
Here, we set the proportion of infected arrivals to 1\% and the proportion vaccinated to 100\%. Additionally, it is assumed that infected arrivals are either pre-symptomatic or asymptomatic. If an arrival is pre-symptomatic their time since infection is sampled at random from the individual's incubation period. If a traveller is asymptomatic, their time since infection on arrival is sampled from the sum of their incubation period and post-incubation period. 

\subsubsection{Filter layer}
{\bf Travellers:} Quarantined travellers are structured into groups who are quarantined together (e.g., in the same room). For the results presented here, the size of these groups is set to 4, and the number of travellers in the facility is set to 100. Travellers are discharged from the system in groups. Once all travellers in a group meet the discharge criteria, the group is removed and a new group generated to replace them. 

{\bf Workforce:} The quarantine workforce is also represented. The workforce model specifies a weekly work schedule for each individual, including 5 days working and 2 days off. On days and off days are scheduled randomly, and a work schedule configuration is accepted by the model as long as at least 5/20 workers are present each day of the week. 

{\bf Transmission control:} We assume that the rate of transmission between travellers who belong to different groups is reduced by a factor of 100, compared to that between travellers who belong to the same group. Similarly, we assume a factor of 100 reduction in transmission between travellers and workers, and a factor of 10 reduction between workers. 

{\bf Vaccination:} The efficacy of vaccination may be varied to reflect available evidence. As above, we assume that some proportion of the workforce, and some proportion of arriving travellers will have been vaccinated. Vaccine efficacy parameters can be varied to reflect the characteristics of vaccines used in particular source countries of interest. In this work, we assume that all individuals in the quarantine system are vaccinated. We implement vaccination by setting efficacy against infection ($V_I = VE$) and efficacy against onward transmission  $V_T = 0$. This choice is not significant when computing transmission within the quarantine system, because all individuals are assumed vaccinated. However, the components of vaccine efficacy could play a significant role when interpreting the force of infection produced by breach events (see sensitivity analysis below). 

{\bf Testing:} Testing of workers can be scheduled at varying frequencies (e.g., daily, every three days, or weekly) and workers will only be tested on days that they attend the workplace. In the results reported here, testing of the workforce is performed daily, as long as a worker is present. 

The model includes testing of travellers that can be scheduled to occur on given days during their quarantine period. For the results reported here, travellers were tested on days 3 and 12 after entering quarantine. We calibrated test sensitivity as specified above, to represent detection via RT-PCR. 

{\bf Response to a positive test or symptoms:} If a worker tests positive or develops symptoms, we assume that they are removed from the workforce and replaced by a new worker (irrespective of whether they attend work that day or not).

If a traveller tests positive or develops symptoms, they are isolated and removed from the quarantine facility (e.g., to a ``health hotel" or hospital), the isolation period is set to 10 days, after which they are released. A positive test result in a traveller triggers a 14-day quarantine extension for other travellers in the same group, and resets their testing schedule (tests performed on days 3 and 12 after extension). 

\subsubsection{Output} 
The output of the quarantine simulation is a time series of ``breach events" with the corresponding characteristics of each individual that leaves quarantine while infected. Several properties are recorded, and the recorded properties depend on whether the infected individual was a worker or traveller. For travellers, the recorded properties are: 
\begin{itemize}
    \item {{\it exposure days}: the number of days they will remain infectious}
    \item{{\it days in quarantine}: the total number of days spent in the quarantine system}
    \item{{\it days in extended quarantine}: the number of days in quarantine extension after detection of a case in a close contact}
    \item{{\it days in isolation}: the number of days spent in isolation after testing positive or presenting symptoms}
    \item {{\it incubation period}: the period between infection and symptom onset}
    \item {{\it time post-incubation}: the period between symptom onset and recovery}
    \item{{\it time discharged}: the timepoint at which the individual was released from quarantine}
    \item{{\it index case (boolean)}: a flag indicating whether the individual arrived infected or was infected while in quarantine}
    \item{{\it symptomatic (boolean)}: a flag indicating whether the individual would express clinical symptoms after their incubation period}
    \item{{\it vaccinated (boolean)}: a flag indicting the individual's vaccination status}
    \item {{\it $\beta_{max}$}: the maximum force of infection for the individual (i.e., force of infection at symptom onset)}
    \item{{\it $\beta_{community}$}: the integrated force of infection over the period after the individual was discharged from quarantine (these are summed to produce the $\beta_{tot}$ value for a simulation)}
\end{itemize}

For workers, the recorded properties are: 
\begin{itemize}
    \item {\it exposure days}: the number of days between infection and detection or recovery
    \item {\it incubation period}: same as for travellers (see above)
    \item {\it post-incubation period}: same as for travellers (see above)
    \item {\it tested positive (boolean)}: a flag indicating whether the individual was removed after testing positive
    \item{\it expressed symptoms (boolean)}: a flag indicating whether the individual was removed after expressing symptoms
    \item {\it time discharged}: the timepoint at which the individual was removed from the facility
    \item{\it symptomatic (boolean)}: same as for travellers (see above)
    \item {\it vaccinated (boolean)}: same as for travellers (see above)
    \item  {\it $\beta_{max}$}: same as for travellers (see above)
    \item {\it $\beta_{community}$}: the force of infection integrated over the period from infection to detection and discharge. 
\end{itemize}

This output timeseries is used to generate input statistics for the index cases used in the branching process model.

\section{Sensitivity analysis of incubation period}
To estimate the effect of a shorter incubation period, we used the statistics reported by the China CDC for an outbreak of the Delta variant in May, 2021 \cite{zhang2021transmission}. While their estimates of the incubation period may differ from those of others \cite{kang2021transmission}, they represent a lower bound for this important parameter. Here, we demonstrate how a shorter incubation period affects the transmission and detection of the virus in our model system. The main differences are:
\begin{itemize}
    \item {Without intervention, slightly less transmission occurs prior to symptom onset.}
    \item {Due to a more rapid increase in viral load, detection of cases typically occurs earlier.}
    \item {In the quarantine model scenario, a shorter overall duration of infection increases the efficacy of a 14-day quarantine.}
\end{itemize}
These results demonstrate how shorter incubation periods can correspond to an enhanced capacity to control transmission. Our results caution against the use of these preliminary estimates of incubation period statistics for use in models of border quarantine systems. As we demonstrate, they potentially over-estimate the effectiveness of these systems and do not represent conservative assumptions about viral dynamics.  

\begin{figure}
    \centering
    \includegraphics[width = 0.5\textwidth]{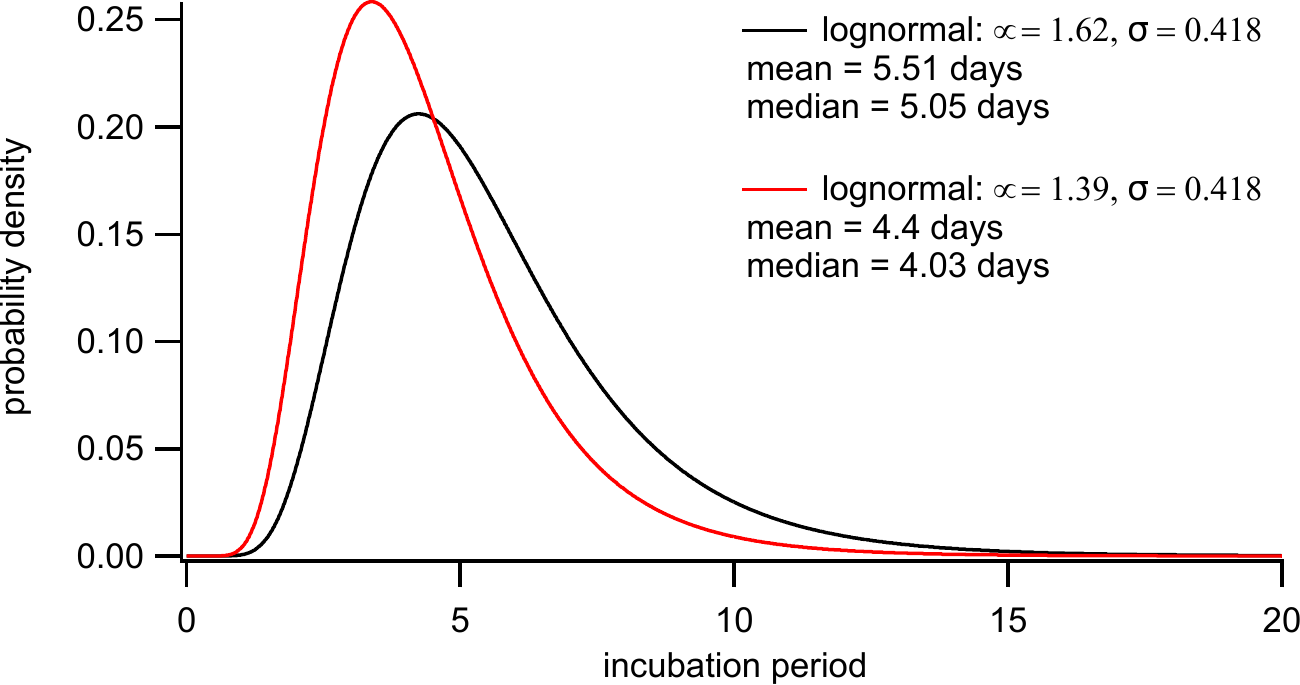}
    \caption{Incubation period distributions used in the Agent-based model. The black trace indicates the distribution used for the main results, while the red trace indicates the distribution used in the sensitivity analysis.}
    \label{fig:tinc_compare}
\end{figure}

\begin{figure}
    \centering
    \includegraphics[width = 0.5\textwidth]{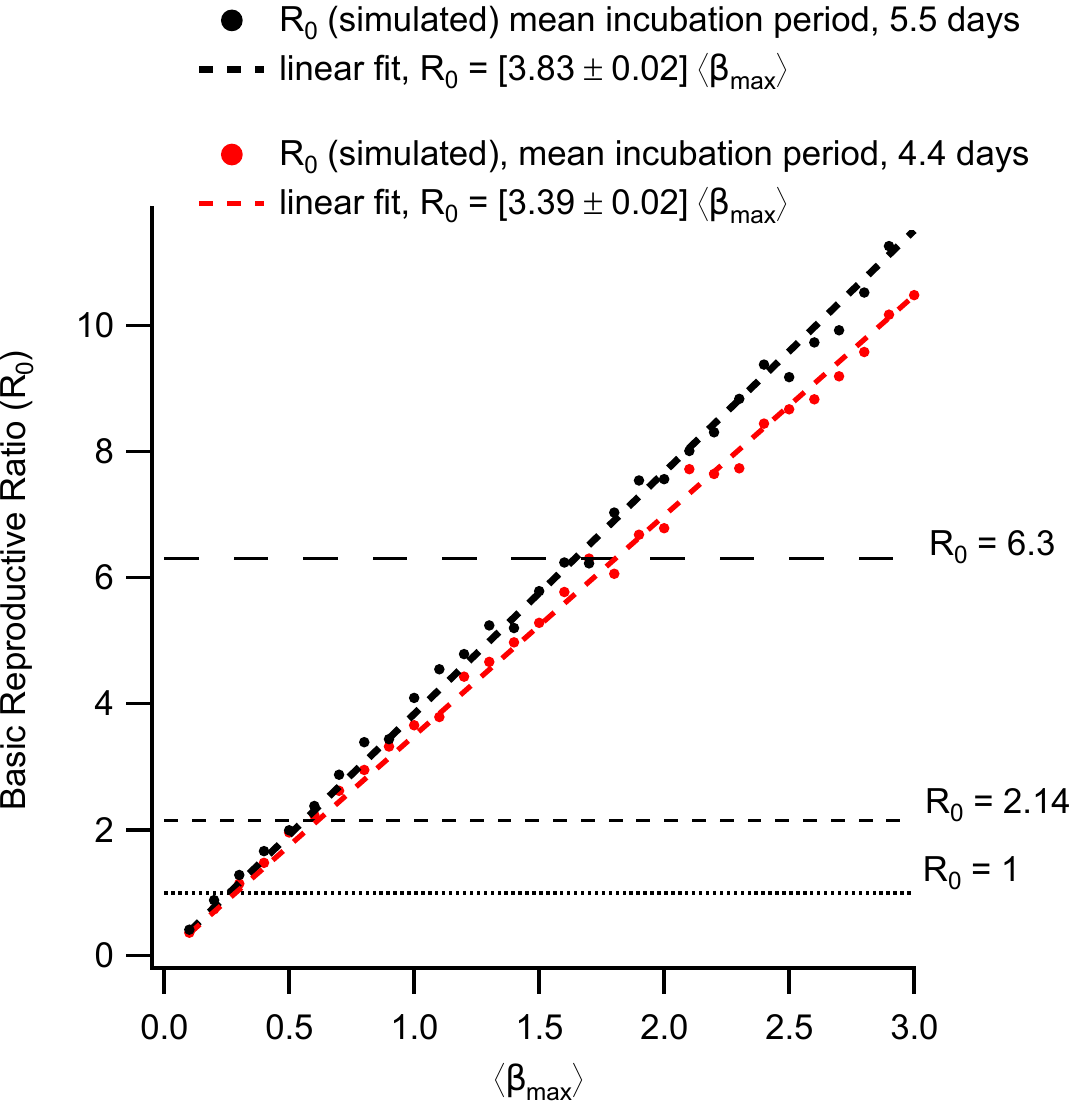}
    \caption{Calibration of $R_0$ as a function of the global transmission scalar for two different incubation periods. }
    \label{fig:tinc_R0_calib}
\end{figure}

\begin{figure}
    \centering
    \includegraphics[width = 0.9\textwidth]{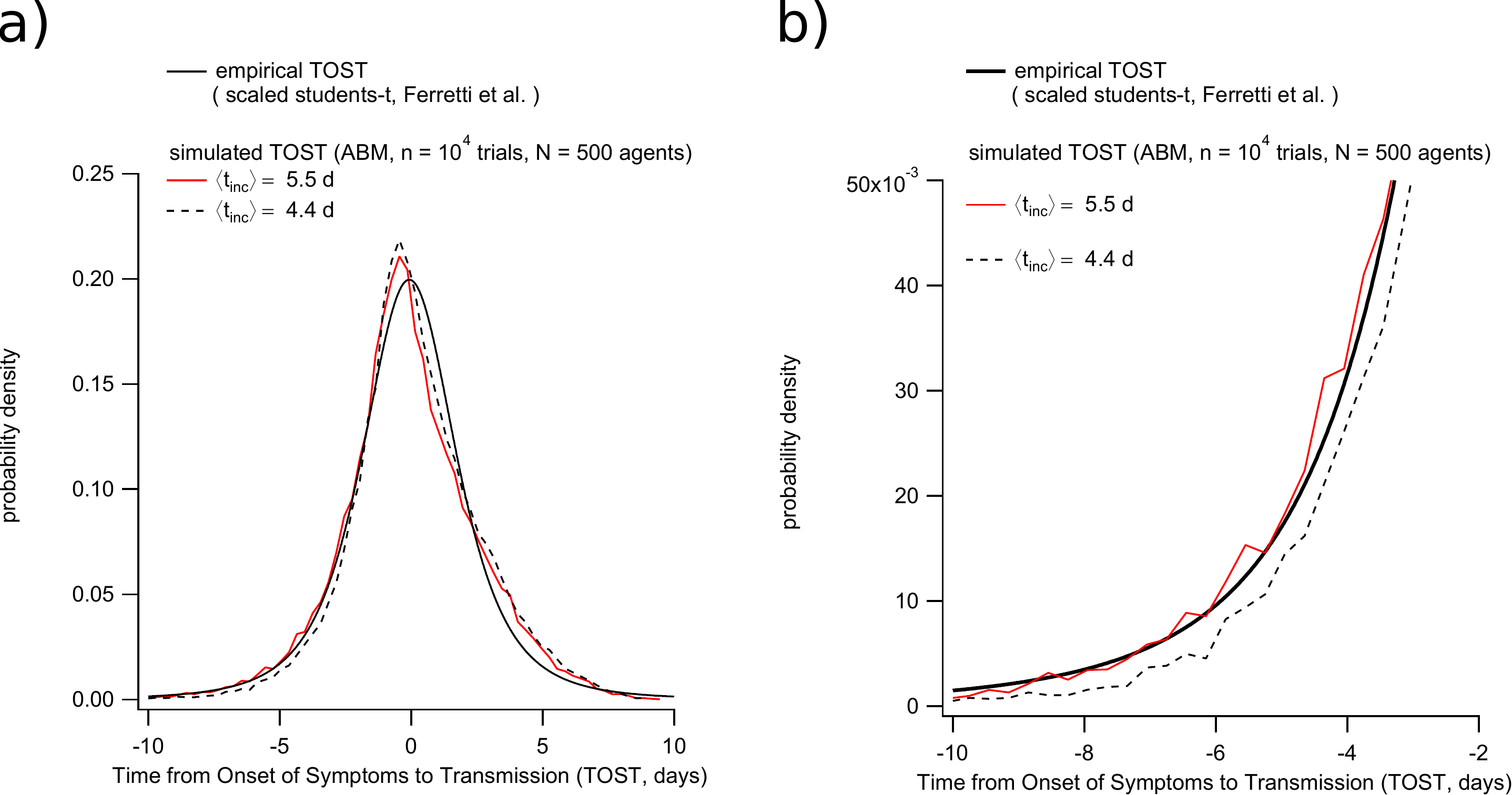}
    \caption{Distribution of TOST for two different incubation period distributions. (a) Shows the full distribution, while (b) zooms in on the negative intervals corresponding to pre-symptomatic transmission.}
    \label{fig:tinc_TOST_compare}
\end{figure}

\begin{figure}
    \centering
    \includegraphics[width = 0.8\textwidth]{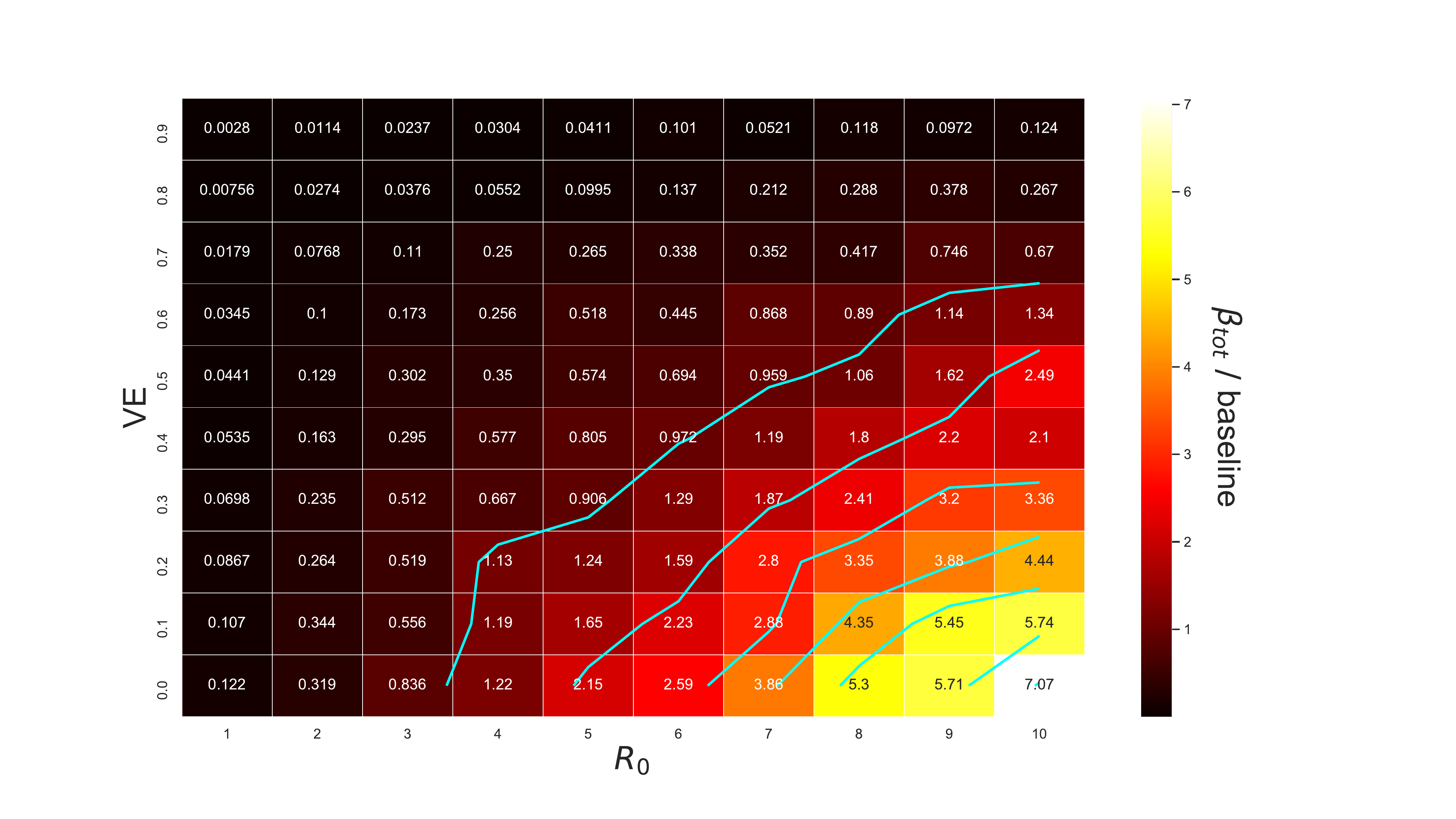}
    \caption{Integrated force of infection ($\beta_{tot}$ relative to baseline) using an incubation period with a mean of 4.4 days. The baseline (denominator) value is taken from the alternate scenario with an incubation period of 5.5 days ($R_0 = 3, VE = 0$). Results are shown as a function of vaccine efficacy and $R_0$.}
    \label{fig:scan_R0xVE_tinc4p4}
\end{figure}

To investigate the effect of incubation period, we first re-calibrated the base model using the modified incubation period distribution (Figure \ref{fig:tinc_R0_calib}). During calibration, we determined the timing of symptom onset relative to transmission, and the timing of detection relative to symptom onset, given daily testing via RT-PCR in a homogeneous transmission network of 500 individuals (Figure \ref{fig:tinc_TOST_compare}). We then applied this modified calibration in the full quarantine simulation to examine the effectiveness of quarantine over the $VE~\times~R_0$ parameter space. The results are qualitatively similar to those produced by the original calibration (average incubation period 5.2 days), but show consistently lower levels of breach risk as measured by $\beta_{tot}$. 

\section{Sensitivity analysis of vaccine efficacy against onward transmission}

\begin{figure}
    \centering
    \includegraphics[width = 0.8\textwidth]{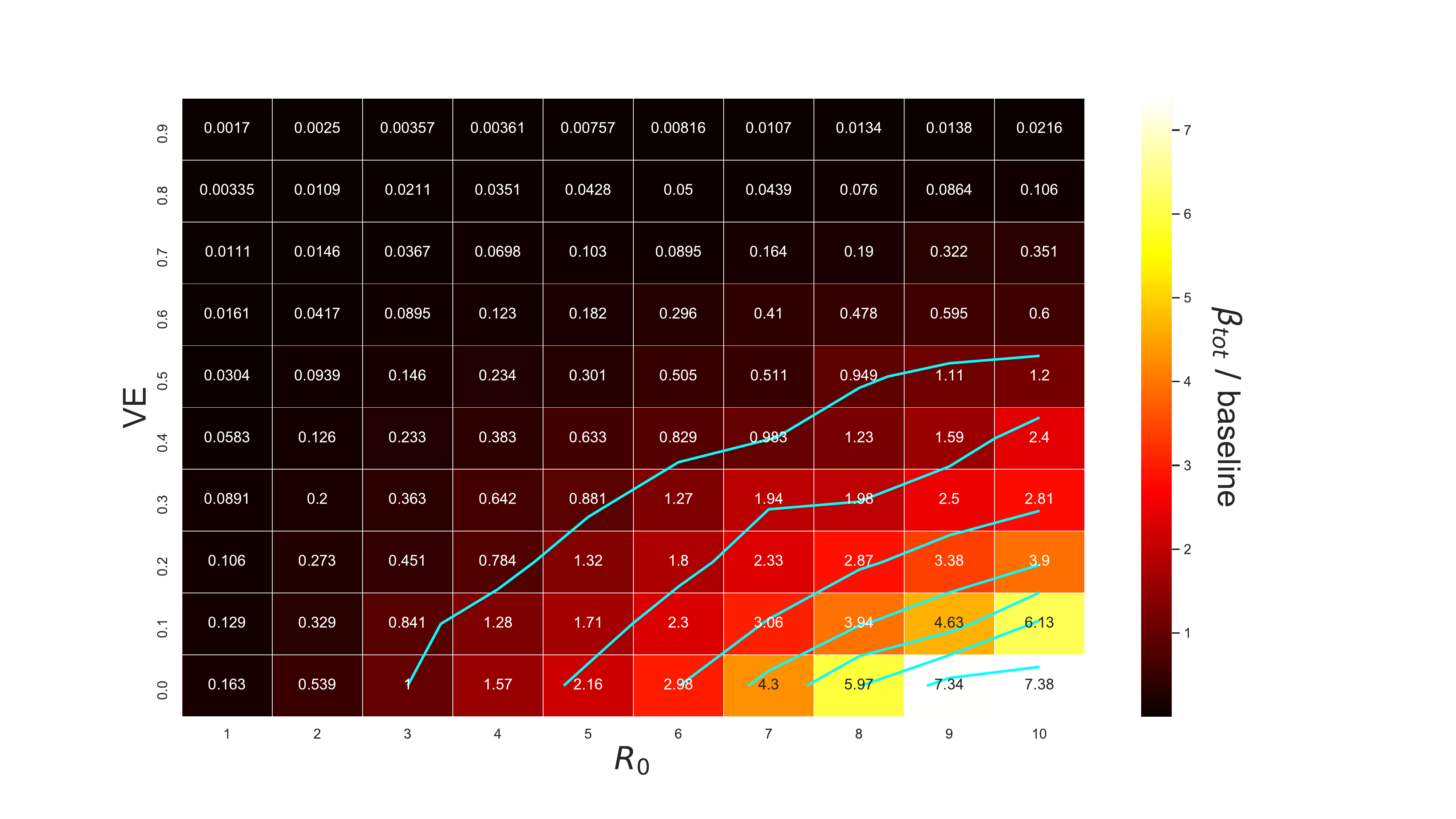}
    \caption{Integrated force of infection ($\beta_{tot}$ relative to baseline) assuming vaccine efficacy against onward transmission is maximised ($V_T = VE$, $V_I = 0$). The baseline (denominator) value is taken from the scenario with $R_0 = 3$, and $VE = 0$. Results are shown as a function of vaccine efficacy and $R_0$.}
    \label{fig:scan_R0xVE_VEimax}
\end{figure}

For the results presented in Figure \ref{fig:net_FoI_vs_R0_x_VE}, it was assumed that vaccine efficacy against onward transmission is negligible. That is, the values reported for $\beta_{tot}$ represent differences in breach statistics rather than the direct effects of vaccine efficacy. Here, we examine the alternate extreme case in which efficacy against transmission is equivalent to $VE$. This equates to multiplication of the $\beta_{tot}$ values in Figure \ref{fig:net_FoI_vs_R0_x_VE} by the factor $1 - VE$. Doing so results in a correction applied to the force of infection values, which fall linearly as $VE$ increases. Under this alternate assumption, for each value of $R_0$ the vaccine efficacy required to maintain baseline outbreak statistics is lower (\ref{fig:scan_R0xVE_VEimax}). The magnitude of the difference increases with $R_0$, ranging from approximately 10\% (for $R_0$ = 4) up to approximately 20\% (for $R_0$ = 10). While these differences are substantial, they represent the largest possible deviation due to efficacy against transmission. Therefore, the results presented in Figure \ref{fig:net_FoI_vs_R0_x_VE} may overestimate the vaccine efficacy required to maintain baseline risk ratios, but the magnitude of this overestimate cannot exceed 20\%. 

\end{document}